\documentclass[acmsmall]{acmart}

\usepackage{subcaption}
\usepackage{threeparttable}
\usepackage{array}
\usepackage{multirow}
\usepackage{algpseudocode}

\usepackage{url}
\usepackage{color}
\usepackage{dsfont}
\usepackage{amsmath}
\usepackage{comment}
\usepackage{setspace}
\usepackage{mhchem,scalerel}
\usepackage{colortbl}
\usepackage{tikz}

\usepackage{algpseudocode}
\usepackage[linesnumbered,ruled,vlined]{algorithm2e}
\SetKwInput{KwInput}{Input}                
\SetKwInput{KwOutput}{Output}              
\newcommand{\sect}[1]{\textsection#1}

\AtBeginDocument{%
  \providecommand\BibTeX{{%
    \normalfont B\kern-0.5em{\scshape i\kern-0.25em b}\kern-0.8em\TeX}}}

\setcopyright{acmcopyright}
\copyrightyear{2020}
\acmYear{2020}
\acmDOI{xxx}

\acmJournal{TOSN}
\acmVolume{1}
\acmNumber{1}
\acmArticle{1}
\acmMonth{11}

\begin{document}

\title{Deep Reinforcement Learning for Tropical Air Free-Cooled Data Center Control}

\author{Duc Van Le}
\affiliation{%
  \institution{Computer Science and Engineering, Nanyang Technological University}
    \streetaddress{50 Nanyang Avenue}
  \country{Singapore}
  \postcode{639798}
}
\email{vdle@ntu.edu.sg}

\author{Rongrong Wang}
\affiliation{%
  \institution{Computer Science and Engineering, Nanyang Technological University}
    \streetaddress{50 Nanyang Avenue}
  \country{Singapore}
  \postcode{639798}
}
\email{rrwang@ntu.edu.sg}

\author{Yingbo Liu}
\affiliation{%
  \institution{Computer Science and Engineering, Nanyang Technological University}
    \streetaddress{50 Nanyang Avenue}
  \country{Singapore}
  \postcode{639798}
}
\email{liuyb@ynufe.edu.cn}

\author{Rui Tan}
\affiliation{%
  \institution{Computer Science and Engineering, Nanyang Technological University}
    \streetaddress{50 Nanyang Avenue}
  \country{Singapore}
  \postcode{639798}
}
\email{tanrui@ntu.edu.sg}

\author{Yew-Wah Wong}
\affiliation{%
  \institution{Energy Research Institute, Nanyang Technological University}
    \streetaddress{50 Nanyang Avenue}
  \country{Singapore}
  \postcode{639798}
}
\email{mywwong@ntu.edu.sg}

\author{Yonggang Wen}
\affiliation{%
  \institution{Computer Science and Engineering, Nanyang Technological University}
  \streetaddress{50 Nanyang Avenue}
  \country{Singapore}
  \postcode{639798}
}
\email{ygwen@ntu.edu.sg}


\thanks{A preliminary version of this work appeared in The 6th ACM International Conference on
Systems for Energy-Efficient Buildings, Cities, and Transportation (BuildSys'19) held in New York, USA, November 2019}

\begin{abstract}
Air free-cooled data centers (DCs) have not existed in the tropical zone due to the unique challenges of year-round high ambient temperature and relative humidity (RH). The increasing availability of servers that can tolerate higher temperatures and RH due to the regulatory bodies' prompts to raise DC temperature setpoints sheds light upon the feasibility of air free-cooled DCs in tropics. However, due to the complex psychrometric dynamics, operating the air free-cooled DC in tropics generally requires adaptive control of supply air condition to maintain the computing performance and reliability of the servers. This paper studies the problem of controlling the supply air temperature and RH in a free-cooled tropical DC below certain thresholds. To achieve the goal, we formulate the control problem as Markov decision processes and apply deep reinforcement learning (DRL) to learn the control policy that minimizes the cooling energy while satisfying the requirements on the supply air temperature and RH. We also develop a constrained DRL solution for performance improvements. Extensive evaluation based on real data traces collected from an air free-cooled testbed and comparisons among the unconstrained and constrained DRL approaches as well as two other baseline approaches show the superior performance of our proposed solutions.
\end{abstract}

\ccsdesc[500]{Hardware~Enterprise level and data centers power issues}
\ccsdesc[300]{Computing methodologies~Reinforcement learning}

\keywords{Data centers, air free cooling, deep reinforcement learning}

\maketitle

\section{Introduction}
\label{sec:intro}

Free cooling that uses the outside air to cool the servers
has emerged as a promising approach to improve the energy efficiency of data centers (DCs)~\cite{cooling-review}.
Free cooling reduces the use of traditional refrigerant-based cooling components such as chillers and compressors. In certain climates, free cooling can save more than 70\% in annual cooling energy of DCs,
which corresponds to a reduction of over 15\% in annualized power usage effectiveness (PUE)~\cite{John:2012}. For instance, a Facebook's air free-cooled DC in Prineville, Oregon reported an annualized PUE of 1.07~\cite{Park:2011} whereas typical DCs have average PUEs of 1.7~\cite{Kevin:2015}.

DCs in the tropical climate with year-round high ambient temperature and relative humidity (RH) consume excessive energy in cooling. However, free cooling in tropics has been long thought infeasible.
For instance, in the target tropic of this paper, the year-round average temperature is about 27\textdegree{}C with record instant maximum of 37\textdegree{}C; the average RH is about 70\% with instant RH up to nearly 100\% before/during rainfalls. If the servers cannot tolerate such high temperatures and RHs, the opportunity of utilizing outside air to cool servers will be very limited.
Fortunately, to prompt DC operators to raise the temperature setpoints for better energy efficiency, the American Society of Heating, Refrigeration and Air-Conditioning Engineers (ASHRAE) has been working on extending the recommended allowable temperature and RH ranges of servers~\cite{Rath:2011}.
For instance, the servers that are compliant with 2011 ASHRAE Class A3 requirement \cite{ASHRAE2011} should be able to operate continuously and reliably with supply air temperature and RH up to 40\textdegree{}C and 90\%, respectively. Many latest servers (e.g., all Dell's gen14 servers and all HPE's DLx gen9 servers) are compliant with the A3 requirement. Such wide allowable ranges for temperature and RH shed light upon the feasibility of air free cooling in tropics. However, ASHRAE's relaxed requirements are for traditional DCs with clean air that is recirculated within the enclosed DC buildings only. In tropics, the free cooling that continuously passes outside air through the server rooms will introduce extra challenges.

An immediate concern is the servers' potential computing performance throttling due to the high supply air temperature. To address this concern, we have conducted extensive measurements on a free-cooled DC testbed that allows us to maintain the supply air temperature in the range of [20\textdegree{}C, 37\textdegree{}C] through a cooling coil and an air heater. The testbed has a total of eight server racks and more than 200 measurement points to closely monitor the state of the testbed including the server room condition and server statuses.
Our 18-month measurements with controlled server room condition and server workload in wide ranges show that the computing performance of the tested servers from four major manufacturers does not drop when the supply air temperature is up to 37\textdegree{}C
(i.e., the record instant maximum in our region). Our measurements show that it is possible to apply air free cooling in tropics without degrading the servers' computing performance.

A major and challenging task in operating air free-cooled DCs in tropics is controlling the condition of the air supplied to the servers, which is the primary focus of this paper. Different from the traditional DCs that use filtrated and circulating air in the enclosed DC building to cool servers, air free-cooled DCs continuously inhale outside air that may contain corrosive gaseous and particulate contaminants. As these contaminants have deliquescent RHs (e.g., 65\%) lower than the tropic's RH, they will absorb the moisture in the air to form corrosive pastes and acids that will undermine the servers' hardware reliability \cite{ASHRAE_Contamination}. Extracting the contaminants from the continuously inhaled outside air will increase capital expenditure (Capex) and operating expenditure (Opex), offsetting or even negating the benefit of air free cooling. To address this challenge, we adopt an approach of mixing a controlled portion of the return hot air from the servers with the fresh outside air to form warm air that will be supplied to the servers. From psychrometrics, the RH of the warm supply air will be lower than that of the fresh outside air. With proper control of the air flows, warm supply air with RH always below the deliquescent RH of the contaminants is beneficial to the server hardware reliability. This RH control approach exploits the heat generated by the servers and their relaxed temperature requirement. When the outside air is too hot, the temperature of the mixed air to meet the RH requirement may exceed the servers' allowable range. In this case, the cooling coil should be used to cool the inhaled outside air.

We formally formulate the control problem as a constrained Markov decision process (CMDP) with the main objective of minimizing the expected energy consumption of the server room fans and cooling coil while  maintaining the supply air temperature and RH below desired thresholds. The control inputs include the supply air volume flow rate, the portion of the return hot air to be mixed with the fresh outside air,
and the temperature drop achieved by the cooling coil when being used.
A key challenge in solving this CMDP problem is the complex psychrometric dynamics.
Specifically, there is no closed-form model to describe the supply air temperature and RH.
In addition, the power consumption of the server room fans, cooling coil and IT equipment
can have complex coupling with the system's psychrometric state.
To address these challenges, we apply deep reinforcement learning (DRL) to learn the optimal control policy over a long time horizon. The DRL control agent iteratively interacts with the air free-cooled DC environment to capture the system dynamics and approaches the optimal control policy.

Moreover, to avoid potential excursions causing thermal unsafety during the online learning phase of DRL's typical workflow, we perform offline learning based on computational models characterizing the psychrometric dynamics and ventilation/cooling energy consumption.
More specifically, we perform psychometric analysis and build two neural networks
for modeling the state evolution of the supply air temperature and RH in the air free-cooled DCs.
Two other neural networks are developed to model the power consumption of the fans and IT equipment.
All developed computational models are validated using the real data traces collected from the air free-cooled testbed mentioned earlier.
The adequately trained DRL agent is then commissioned to control the air free-cooled DC.
We extensively evaluate the performance of the proposed unconstrained and constrained DRL-based control approaches based on real data traces collected from the testbed and show its effectiveness through the comparison with two other baseline approaches which are hysteresis-based and model-predictive control approaches.
The evaluation results show that our constrained DRL approach can achieve more cooling energy saving and
less supply air temperature and RH requirement violations under various temperature and RH requirements,
compared with other approaches.

To the best of our knowledge, this is the first work that studies the server room condition control based on DRL techniques for air free-cooled energy-efficient DCs in tropics with year-round high temperature and RH. Our results provide an important basis for full implementations and DRL-based control of air free-cooled DCs on the testbed and in production settings.

The remainder of this paper is organized as follows. \sect\ref{TDC_discussion} introduces the background of free cooling and reviews related research. \sect\ref{measurement} discusses the requirements of air free-cooled DCs in tropics. \sect\ref{sec:problem_formulation} formulates the control problem. \sect\ref{sec:drl} presents the DRL-based control approaches. \sect\ref{sec:eval} presents evaluation results. \sect\ref{sec:conclude} concludes this paper.

\section{Air Free-Cooled DC \& Related Work}
\label{TDC_discussion}

\subsection{Background}
\label{subsec:background}

DCs generally use different types of cooling systems to move the heat generated by the servers, depending on the DC sizes. Large-scale DCs often use a cooling system consisting of {\em water chillers} and {\em computer room air handlers} (CRAHs)~\cite{Cho:2014}. CRAHs circulate the hot return air carrying the heat generated by the servers to the cooling coils which lower the temperature of the return air to a desired setpoint by transferring the heat via the chilled water to the chillers. The hot water carrying the heat at the chillers is then moved in another water loop to the outdoor {\em cooling towers} which transfer a portion of the heat via evaporation to the ambient. Finally, the chillers remove the remaining heat and return the chilled water with a specific temperature setpoint (e.g., 7\textdegree{}C) to the cooling coils.
Differently, small-scale DCs often use direct-expansion compressor-based air conditioners to take away the heat. For both chiller-CRAH systems and compressor-based air conditioners, the air is circulated in the DC building only and not mixed with the outside air. The closed-loop air circulation facilitates precise controls of the temperature, RH, and cleanliness of the air in the server rooms.
However, the water chillers, cooling towers, and compressors are power-intensive facilities. Therefore, these conventional cooling systems in general consume a significant amount of energy.

In recent years, {\em free cooling} has emerged as an effective scheme to improve DCs' energy efficiency~\cite{cooling-review}.
It obviates the need of power-intensive water chillers and compressors by passing outside air through some heat dissipation device.
A recent study in \cite{Gupta:2019}
has shown that by combining free cooling and solar generation, DC's brown energy consumption can be reduced by up to 59\%. There are two major free cooling forms~\cite{Christy:2011}, i.e., water-side and air-side methods.
In the water-side method, an energy-free {\em heat exchanger} uses water to carry the heat
to the outdoor cooling towers. Large fans blow the outside air through the cooling towers to dissipate the heat into the ambient. Differently, the air-side method uses fans to blow the outside air directly into the server rooms without the water intermediary. The hot return air carrying the heat is then guided back using fans into the ambient. To adjust the temperature and/or RH of the air supplied to the server rooms, a portion of the return air can be recirculated and then mixed with the fresh incoming air.
Note that free cooling admits minimum use of traditional cooling systems when the control target for the supply air condition cannot be achieved by purely using the water-side or air-side methods. In this study, we focus on the air-side method (referred to as {\em air free cooling}) due to its simplicity and higher energy efficiency. In particular, we study the control of supply air condition in air free-cooled DCs in the tropical climate that imposes a number of unique challenges as we will discuss in \sect\ref{measurement}.

\subsection{Related Work}
\label{related}

In what follows, we review related work on air free cooling control and various applications of DRL for saving energy.

\subsubsection{Air free cooling control}
A few existing studies \cite{Goiri:2013,Goiri:2015,Manousakis:2015} focus on the supply air condition control in air free-cooled DCs.
Goiri~\emph{et al.}~\cite{Goiri:2013} designed and implemented a real air free-cooled DC testbed called Parasol, which combines the air free cooling with a direct-expansion air conditioner to control the supply air temperature.
The work in~\cite{Goiri:2015} proposed an air free cooling control approach called CoolAir to
maintain the average and variation of supply air temperatures in desired ranges. Based on predicted ambient temperatures, it selects a proper temperature setpoint to limit the temperature variation and minimize the use of traditional cooling.
The work in~\cite{Manousakis:2015} presented an optimization framework that determines the optimal provision from the traditional cooling in an air free-cooled DC to reduce cooling-related Capex and Opex subject to temperature constraints.
The above studies \cite{Goiri:2013,Goiri:2015,Manousakis:2015} focused on the temperature control to avoid server shutdown due to overheating.
RH control is usually not considered because free cooling has been recommended for cold and dry locations only~\cite{Lee:2013}, where the ambient RH does not exceed 60\% in general. Differently, in tropics, the ambient RH is high.
From the study by Manousakis~\emph{et al.}~\cite{Manousakis:2016} based on data collected from a number of Microsoft air free-cooled DCs,
the hard disk drive (HDD) failure rate is 10x more correlated with RH than temperature.
Therefore, in this paper, we jointly address temperature and RH controls in air free-cooled DCs such that the energy savings achieved by the air free cooling can be
maximized while maintaining high server performance and reliability.

\subsubsection{DRL for DC control}
Reinforcement learning (RL)~\cite{Sutton1998} is a trial-and-error learning approach, in which an {\em agent} explores and learns the optimal control policy by interacting with its {\em environment}
through a sequence of the environment's states, the actions applied on the environment, and the rewards.
DRL that uses a deep Q-network (DQN)~\cite{DQN2015} as the function approximation of the control policy
for the agent, has emerged as an effective method for solving complex control problems
with high-dimensional state and action spaces.
DRL has been recently applied to develop control policies for the heating, ventilation, and air-conditioning (HVAC) systems of buildings~\cite{Wei:2017,Zhang:2018}. Wei~\emph{et al.}~\cite{Wei:2017} developed a DRL agent for
HVAC control in the presence of ambient dynamics. The DRL agent chooses the optimal air flow rates for different zones in the building such that the energy consumption is minimized subject to tenants' comfort requirements.
Zhang~\emph{et al.}~\cite{Zhang:2018} implemented and evaluated a practical DQN agent for a radiant heating system that aims to improve a building's energy efficiency. A three-month experiment in \cite{Zhang:2018} shows that the DQN agent resulted in up to 18.2\% heating demand reduction, compared to the rule-based control. Chen~\emph{et al.}~\cite{Chen:2019} developed a DRL agent for HVAC control, called Gnu-RL.
The Gnu-RL is pre-trained offline on historical data using imitation learning to mimic a deployed proportional-integral-derivative (PID) controller. Then, it interacts with the environment during online training to improve its end-to-end control policy. Evaluation results of controlling the amount of air flow supplied to a real conference room show that Gnu-RL can save 16.7\% of cooling demand,
compared with an existing controller using a variable air volume box.

Due to RL's trial-and-error nature, DRL has not been widely used for environment condition control in mission-critical DCs that often have tight requirements on temperatures. Google reported the adoption of DRL for cooling control in several of its DCs \cite{Knight:2018}. However, Google does not release any technical details. Li~\emph{et al.}~\cite{Yuanlong:2017} developed a DRL-based cooling DC control framework for energy cost minimization with supply temperature constraints and achieved about 11\% cooling energy saving in simulations, compared with two baseline schemes.
Yi~\emph{et al.}~\cite{yi2019} applied DRL to allocate computing jobs and reduce servers' energy consumption.
To avoid potential unsafety caused by DRL's trial-and-error,
the DRL training is performed offline using computational models capturing servers' power and thermal dynamics.
Our work applies DRL for controlling air free-cooled DCs in tropics.
DRL well addresses the complex thermal and psychrometric dynamics.
Similar to~\cite{Zhang:2018,yi2019}, we also adopt an offline training approach to preclude
the risk caused by the trial-and-error nature of the learning phase.
While our paper and the existing study~\cite{Yuanlong:2017,yi2019} share the same control objective (i.e., to reduce energy consumption), we address different physical dynamics and constraints of the air free cooling design.

A preliminary version of this work~\cite{le:2019} presented an unconstrained DRL-based approach for controlling the air free-cooled DCs with the main objective of minimizing the cooling energy consumption while satisfying the requirements on the supply air temperature and RH requirement.
The preliminary work merged the main objective and penalties of requirement violations
into a weighted reward function with constant penalty coefficients.
The unconstrained DRL agent is then trained to learn a control policy that maximizes the long-term weighted reward. Due to the usage of the constant penalty coefficients,
the unconstrained DRL agent may converge to a control policy
which minimizes the cooling energy or supply temperature/RH constraint violation penalties only.
Moreover, selection of proper constant coefficients for different supply
air temperature and RH constraint requirements is non-trivial.
In this paper, to address these limitations, we develop a constrained DRL approach which
can adaptively learn the optimal penalty weights based on Lagrangian primal-dual policy optimization.
Our new evaluation results show that our constrained DRL approach achieves more cooling energy saving
and less violations of the supply air temperature and RH requirements, compared with our prior unconstrained DRL and other baseline approaches.

\section{Air Free-Cooled DC in Tropics}
\label{measurement}

In this section, we present the design of an air free-cooled DC testbed located in the tropical zone (\sect\ref{subsec:testbed}). Then, we discuss the temperature requirement (\sect\ref{subsec:temp-requirement}) and RH requirement (\sect\ref{subsec:RH-requirement}) for operating air free-cooled DCs in tropics.

\subsection{Air Free-Cooled DC Testbed}
\label{subsec:testbed}

To study the feasibility of air free cooling in tropics with high ambient temperature and RH,
we designed and instrumented an air free-cooled DC testbed located in the tropical zone.
The testbed consists of two identical side-by-side server rooms that are located within the premise of a DC operator.
In what follows, we briefly describe the design of a server room. More details can be found in~\cite{tech-report}.

\begin{figure*}
  \centering
  \begin{subfigure}{.5\textwidth}
    \includegraphics[width=\textwidth]{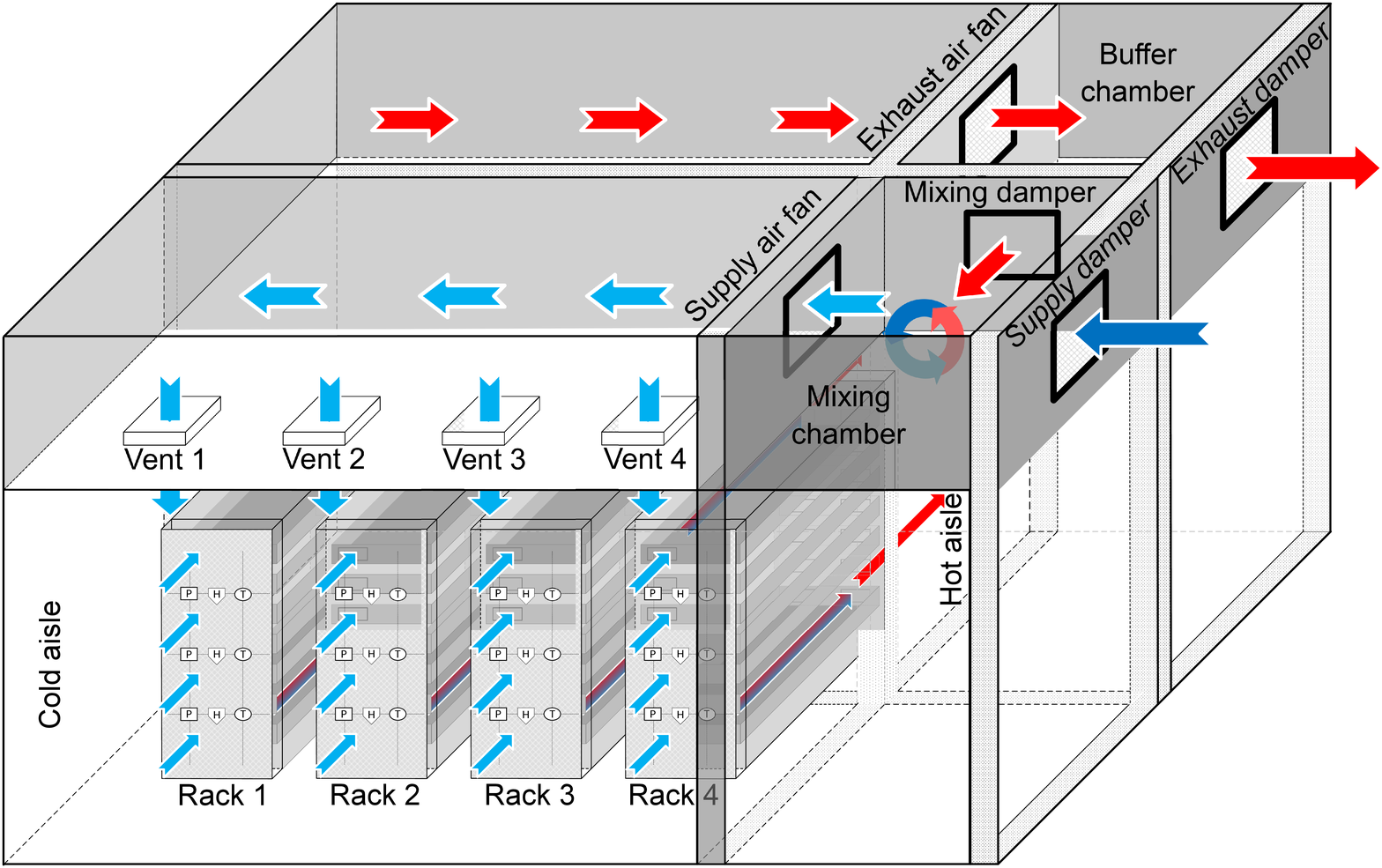}
    \caption{3D view.}
    \label{fig:3d}
  \end{subfigure}
  \begin{subfigure}{.245\textwidth}
    \includegraphics[width=\textwidth]{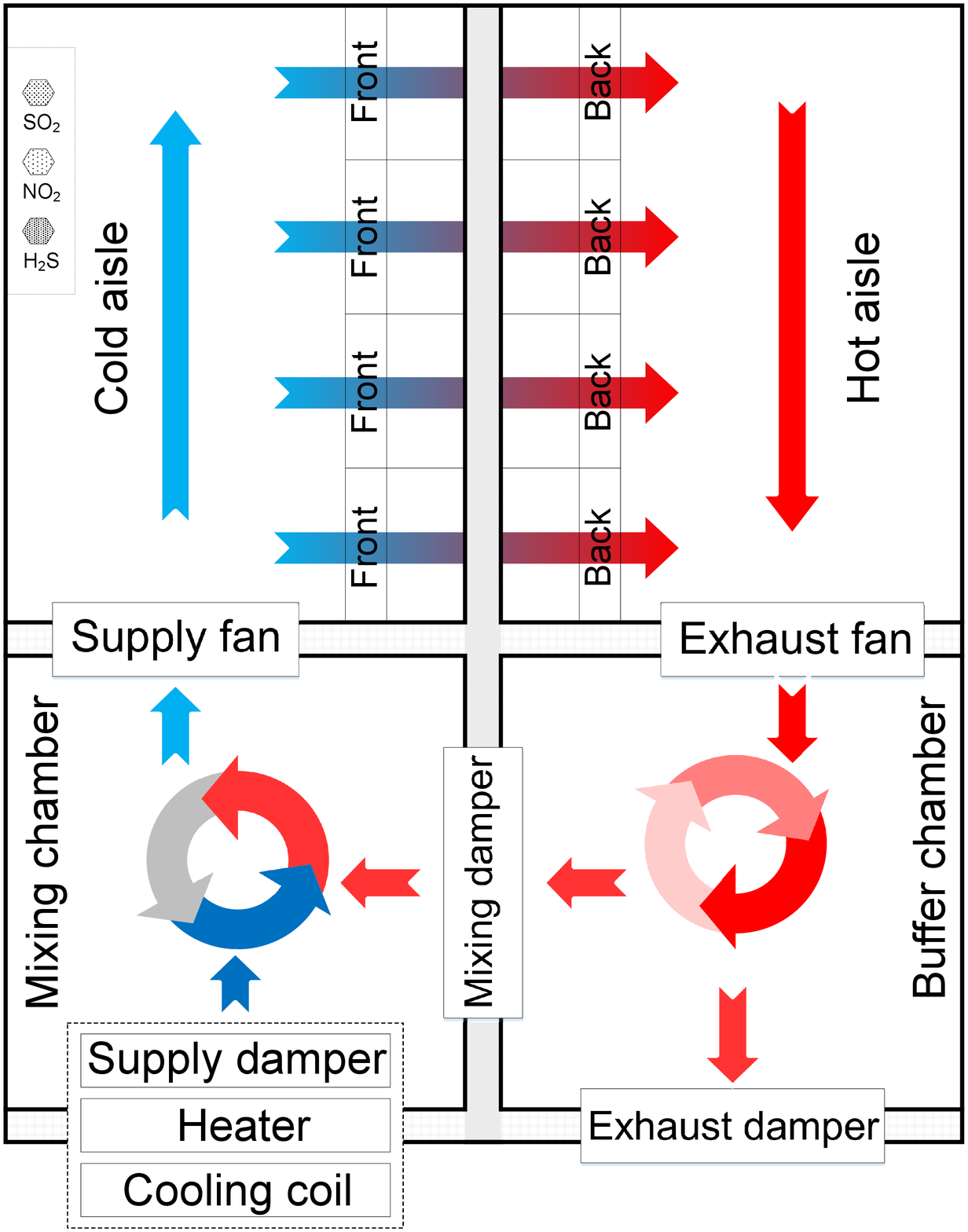}
    \caption{Top view.}
    \label{fig:top}
  \end{subfigure}
  \hspace{0.02\textwidth}
  \begin{subfigure}{.21\textwidth}
    \includegraphics[width=\textwidth]{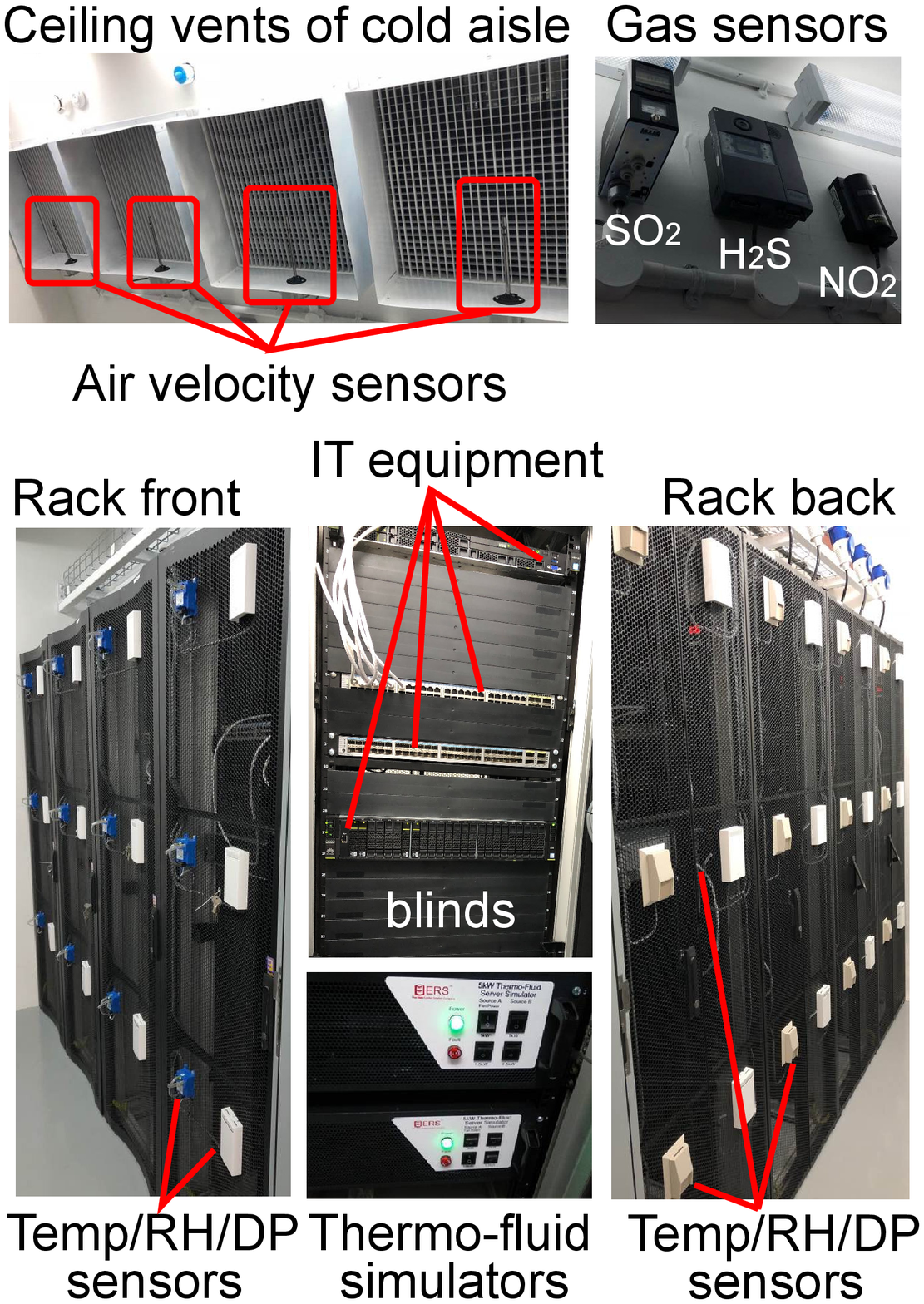}
    \caption{Equipment.}
    \label{fig:equipment}
  \end{subfigure}
  \caption{The air free-cooled DC testbed used in this work. Arrows represent the air flows.}
  \label{fig:room}
\end{figure*}

Figs.~\ref{fig:3d} and \ref{fig:top} show the 3D and top views of the server room, respectively. The room has two layers with each divided into four chambers. A cooling coil and an air heater are installed on the top layer to process the fresh air inhaled into the test room. Note that the air heater is used only in a set of tests investigating the performance of the servers in high temperatures (cf.~\sect\ref{subsubsec:impact-performance}). The air free cooling control does not use the heater. Two fans (i.e., supply fan and exhaust fan) are installed on the top layer to move air. Moreover, there are three dampers (i.e., supply damper, exhaust damper, and mixing damper) as shown in Fig.~\ref{fig:room}. By setting their openness, we can control the air flow paths. The three dampers together are referred to as {\em damper system}. After the supply fan, the air enters a chamber and then goes down to the cold aisle chamber on the bottom layer through four vents. This design improves the evenness of the cold air volumes passing through the vents. Four 42U server racks are installed on the bottom layer, sitting between the cold aisle and hot aisle chambers. Our design well separates the cold air supplied to the servers and the hot air generated by them. This facilitates the control of the condition of the air supplied to the servers. The hot air is moved by the exhaust fan into a buffer chamber. Depending on the damper system's setting, the hot air is exhausted and/or recirculated to the mixing chamber.

Fig.~\ref{fig:equipment} shows the deployment of some IT equipment and sensors. The racks in each server room host a total of six servers and five 1Gbps switches made by four different manufacturers. All these IT devices were new when they were deployed. To generate more heat and improve the realism of the testbed, for each server room, a total of six thermo-fluid server simulators are mounted on the racks. Their power consumption can be configured and can reach $30\,\text{kW}$ totally that is comparable to that of about 100 servers. To well separate the cold and hot aisles, we deploy blinds for the rack slots not mounted with IT equipment and thermo-fluid server simulators.
We also install a total of 85 sensors of various modality in each room to monitor the environmental condition as well as the powers consumed by the room facility and IT equipment.
Specifically, we deploy the following sensors: (1) a combined temperature and RH sensor outside of the server room to monitor the ambient condition; (2) a temperature sensor in each of the mixing, cold aisle, hot aisle, and buffer chambers; (3) an air velocity sensor at each of the four cold vents to estimate the air volume speed in $\text{m}^3/\text{h}$; (4) sensors in the cold aisle for monitoring differential pressure (DP) with respect to atmospheric pressure and concentrations of corrosive gases (SO$_2$, NO$_2$, H$_2$S); (5) temperature, RH, and DP sensors at three heights on the front and back sides of each rack;
(6) power meters to monitor the power of server rack, cooling coil, heater, and fans.
The dense sensor deployment is for research only. In \sect\ref{subsec:requirement}, necessary sensors for free cooling control will be discussed.

The real-time measurements of several sensors (e.g., temperature and air volume speed) are also used by various control algorithms to maintain the test room's environmental condition. For instance, the total air volume speed supplied to the servers can be maintained at a specified setpoint up to $12500\,\text{m}^3/\text{h}$ by a PID controller for the supply and exhaust fans.

\subsection{Supply Air Temperature Requirement}
\label{subsec:temp-requirement}

\subsubsection{Impact of temperature on server safety and reliability}

Too high instantaneous supply air temperatures may cause permanent damages to server hardware components. To avoid the damage, most servers will automatically halt for self protection when the temperatures measured by the built-in sensors of the server enclosure exceed certain {\em safety thresholds}. For instance, a server deployed on our testbed has a safety threshold of 45\textdegree{}C for its inlet temperature sensor. For continuous operation of a DC, the safety thresholds of the servers must not be exceeded.

Besides the permanent damages caused by too high temperatures instantly, high temperatures are generally thought generating negative impact on the server hardware's long-term reliability that is often measured with annualized failure rate (AFR). A basis of this hypothesis is the Arrhenius equation that characterizes the temperature dependence of reaction rates in physical chemistry \cite{laidler1965chemical}. The electronics industry adopts this equation to predict that the failure rate of an electronic device increases exponentially with the temperature \cite{arrhenius}. Based on this, ASHRAE, together with DC IT equipment manufacturers, provides the {\em x-factors}, which are the relative failure rates under certain temperatures, as a guideline for choosing DC temperature setpoint \cite{ASHRAE2011}. For instance, with a temperature of 37.5\textdegree{}C, the x-factor is 1.61, meaning that the failure rate at 37.5\textdegree{}C will be 1.61 times of the failure rate at the reference temperature of 20\textdegree{}C. For example, if the baseline AFR of HDDs at 20\textdegree{}C is 1.25\% according to a cloud service provider's statistics \cite{backblaze}, the AFR at 37.5\textdegree{}C is $1.25\% \times 1.61 \simeq 2\%$, i.e., two out of 100 HDDs fail over one year. Since the baseline AFR for any server component is low in general, the absolute increases of AFR due to higher temperatures are not significant.
In particular, the recent advances in materials development and hardware design enable manufacturers to build more robust DC IT equipment that can tolerate higher temperatures and RHs. For example, many modern servers (e.g., all Dell's gen14 servers and all HPE's DLx gen9 servers) are compliant with ASHRAE Class A3 requirement~\cite{ASHRAE2011}. Specifically, these servers can continuously and reliably operate under a temperature range of $[5\text{\textdegree{}C}, 40\text{\textdegree{}C}]$ and RH range of $[8\%, 85\%]$. We call the temperature upper limit for a server's design reliability as {\em reliability threshold}, e.g., 40\textdegree{}C for ASHRAE Class A3 servers. Note that for a server, the reliability threshold is in general lower than the safety threshold, because the latter concerns about instant damages.

The ambient temperature of the tropical area that we are in has a record minimum of 19.4\textdegree{}C and maximum of 37.0\textdegree{}C. Thus, by using air free cooling only, it is possible to maintain the supply air temperature below modern servers' reliability thresholds. However, close monitoring and cautious control of the supply air temperature are still needed, because of the following. First, uncontrolled hot air recirculation due to imperfect separation of the cold air and hot air aisles may increase the supply air temperature. Second, as discussed shortly in \sect\ref{subsec:RH-requirement}, to reduce RH, an energy-efficient approach is to use controlled hot air recirculation to raise the supply air temperature. However, it reduces the buffer region from the safety thresholds. Thus, without cautious control, the system will have increased risk of server shutdown caused by overheating.

\subsubsection{Impact of temperature on server performance}
\label{subsubsec:impact-performance}

Another common concern is that high temperatures may cause degraded computing performance of servers. We conduct extensive controlled experiments over a duration of about eight months to investigate the impact of supply air condition (temperature and volume flow rate) on the server performance. We concluded that, when the supply air temperature is up to 37\textdegree{}C (i.e., the record maximum in our area), the temperature has no impact on the servers' computing performance if a sufficient air flow rate is maintained (e.g., $2,500\,\text{m}^3/\text{h}$ for one server room of our testbed). This section briefly summarizes the experiment methodology and results.
\

\begin{table}
\caption{Settings for server performance benchmark.}
\label{table_parameter}
\centering
\begin{tabular}{lllll}
\toprule
\textbf{Parameter} & \textbf{Minimum} & \textbf{Maximum} & \textbf{Steps} \\
\midrule
Supply air temperature & 25\textdegree{}C & 37\textdegree{}C & 13\\
Room air flow rate ($\text{m}^3/\text{h}$) & 2,500 & 12,500 & 5 \\
CPU utilization & 10\% & 90\% & 7 \\
HDD throughput (MB/s) & 10 & 100 & 6 \\
Memory block size (KB) & 8 & 256 & 6 \\
\bottomrule
\end{tabular}
\end{table}

We separately benchmark the CPUs, HDDs, and main memories, which are the main components related to servers' computing performance. For each component, we vary the supply air temperature, the air volume flow rate, and the operating setpoint of the tested server component in their respective ranges. Table~\ref{table_parameter} summarizes the ranges of these parameters and the corresponding numbers of steps. Under each setting, we conduct a 1-hour experiment to measure giga floating point operations per second (GFLOPS) for CPU, input/output operations per second (IOPS) and response time for HDD, and speed of data coping for memory. Benchmark results for a total of 1,235 net test hours have been collected.
Table~\ref{server_performance} shows the benchmark results for a CPU, an HDD, and a memory, under different supply air temperatures, and specific settings of CPU utilization, HDD throughput, and memory block size.
We can see that the performance metrics remain stable when the temperature is up to 37\textdegree{}C. Other CPUs, HDDs, and memories also exhibit such stable trend. We also conducted experiments to jointly benchmark CPU, HDD, and memory, such that all these components generate heat simultaneously. Similarly, we observed no statistically significant impact of temperature on the computing performance within the test ranges specified in Table~\ref{table_parameter}.
More details of result analysis and observations on our microbenchmarks spanned 18 months using the designed air free-cooled testbed can be found in~\cite{tech-report}.

\subsection{Supply Air RH Requirement}
\label{subsec:RH-requirement}

RH is the ratio of the amount of moisture contained in the air at a given temperature to the maximum amount of moisture that the air can hold at the same temperature. As discussed in \sect\ref{subsec:temp-requirement}, modern servers can operate reliably under high RHs (up to 85\%) under typical DC settings. In typical air cooled DCs, the air is circulated within the DC building without admitting much fresh air from the outside; any admitted fresh air will be filtrated to control the concentrations of gaseous and particulate contamination \cite{ASHRAE_Contamination}. Research has shown that, with clean air, RH has little impact on the IT hardware reliability \cite{Singh:2015}.

Differently, in the air free cooling scheme, the outside air continuously passes through the server rooms. The solutions to control the concentrations of gaseous and particulate contamination will increase Capex for installing the filtration facility and Opex for filtration energy consumption and consumable component replacement. Thus, the design of our testbed chooses not to integrate the costly continuous air filtration solutions; it only applies a MERV 6 filter to remove PM10 or larger particles. Finer particles and corrosive gases (e.g., \ce{SO_2}, \ce{H_2S}, \ce{NO_2}, and \ce{Cl_2}) generated by transportation systems and industrial processes can negatively affect the reliability of the IT equipment. Specifically, if the RH of the supply air is higher than the deliquescent RH of the particles and gases, these contaminants will absorb the air moisture to form corrosive pastes and acids that will promote corrosion and/or ion migration of the IT hardware materials \cite{ASHRAE_Contamination}. Corrosion can easily cause short circuits given today's dense layouts of printed circuit boards.

\begin{table}[t]
  \caption{Server performance benchmark results.}
\begin{tabular}{ccccc}
\toprule
\textdegree{}C & GFLOPS  & IOPS & RespTime (ms) & MemSpeed (MB/s)\\
\midrule
25 & 303.97 & 6397.0 & 0.28 & 2873.12\\
27 & 301.48 & 6397.0 & 0.25 & 2990.79\\
29 & 282.41 & 6398.0 & 0.25 & 3539.04\\
31 & 300.01 & 6398.0 & 0.27 & 3191.80\\
33 & 308.63 & 6398.0 & 0.24 & 2977.22\\
35 & 295.19 & 6401.0 & 0.32 & 2792.49\\
37 & 304.15 & 6398.0 & 0.25 & 2798.15\\
\bottomrule
\multicolumn{5}{l}{\footnotesize CPU utilization: 90\%; HDD throughput: 100 MB/s; memory block size: 256 KB}
\end{tabular}
\label{server_performance}
\end{table}

Existing studies have shown that the co-presence of high RH and air contaminants lead to reduced server hardware reliability. Svensson~\emph{et al.}~\cite{Svensson:1993} observed that the increase of RH from 75\% to 95\% results in about 9x higher corrosion rate of zinc at the same concentration level of \ce{SO_2}. A study \cite{Singh:2015} showed that, in the presence of gaseous contamination, the copper corrosion rate of DC IT equipment increases with RH.
Therefore, it is important to maintain low RH for the supply air. This is a challenging requirement because the ambient RH in tropics is generally high. From our measurements, the ambient RH has an average of 71\% with instant measurements up to 100\%. Note that the deliquescent RH for many contaminants is about 65\% \cite{ASHRAE_Contamination}.

\section{Problem Formulation}
\label{sec:problem_formulation}

From \sect\ref{subsec:temp-requirement} and \sect\ref{subsec:RH-requirement}, to achieve success of air free-cooled DCs in tropics, we will need to maintain the supply air temperature below the reliability thresholds of the servers and RH below a certain level (e.g., the lowest deliquescent RH of the particulate and gaseous contaminants present in the outside air).
In this section, \sect\ref{subsec:approach-overview} overviews our approach to meeting the requirements; \sect\ref{sec:cmdp} presents a constrained Markov Decision Process (CMDP) formulation of the control problem, which will be addressed by using the DRL in \sect\ref{sec:drl}.

\subsection{Approach Overview}
\label{subsec:approach-overview}

RH control is a challenging task in tropics. Traditionally, dehumidification is achieved by a cooling-then-reheating process. Specifically, a certain amount of moisture is condensed out from the humid air by cooling the air below its dew point. Then, the cold air is reheated to the desired temperature. However, the cooling and reheating processes consume significant energy.
In this study, to reduce the supply air RH in an energy-efficient manner, we recirculate a portion of the hot return air and mix it with the fresh outside air to form supply air. The mixing can be implemented by controlling the openness of the three dampers as illustrated in Fig.~\ref{fig:room}. Note that, without condensation, the hot return air and the fresh outside air have the same absolute humidity. From psychrometrics, the hotter mixed air will have lower RH compared with the fresh outside air.
However, when the fresh outside air is hot, the hotter mixed air to achieve the desired low RH may exceed the servers' reliability thresholds. In this case, the cooling coil should be activated to reduce the temperature of the incoming air.

This paper develops control algorithms for the supply and exhaust fans, the cooling coil, and the dampers such that the energy consumption of the non-IT facility is minimized subject to that
the temperature and RH of the air supplied to the servers are below
respective specified thresholds for the sake of IT hardware reliability. The system will operate in the presence of exogenous disturbances, i.e., the time-varying ambient condition and heat from servers.

\subsection{Control Problem Formulations}
\label{sec:cmdp}

Time is divided into intervals with identical duration of $\rho$ seconds.
In this paper, we consider the secondary controls (i.e., adjustment of setpoints) for the actuators.
The beginning time instant of a time interval is called a {\em time step}.
Control action is performed at every time step.
Thus, the $\rho$ is referred to as {\em control period}.
In this paper, we do not consider the details of the primary controls of the actuators; we assume that the actuators can implement the setpoints decided by the secondary controls using their closed-loop primary controls and the system has reached the steady state by the end of every control period. In practice, the setting of the control period can be chosen with the consideration of the dynamics of the primary controls to ensure the above assumption.
Under the above setting, the temperature and RH of the supply air at next time step depend only on the system's state (conditions of outside and supply air, servers' powers) and the control action at the current time step (cf.~\sect\ref{tdc_modeling}).
Furthermore, the control policy should satisfy the supply air temperature and RH constraints.
Therefore, the control problem can be modeled as Markov decision processes (MDPs).
We now define the terminologies of the MDP formulations.

\subsubsection{System state}
The system state, denoted by $x$, is a vector $x = [t_s, \phi_s, p_{\text{IT}}, t_o, \phi_o]$, where $t$ and $\phi$ respectively represent temperature and RH, the subscript $s$ and $o$ respectively represent supply air and outside air, and $p_{\text{IT}}$ represents the total power consumption of all IT equipment in the server room. The $p_{\text{IT}}$ determines the amount of heat generated in the server room.
Let $X$ denote the set of all possible states in the system.

\subsubsection{Control action}
The supply and exhaust fans admit air volume flow rate setpoints. To achieve steady state without control errors, the setpoints for the two fans should be identical; otherwise, the server room will be in the dynamic process of pressurization/depressurization or a steady state with control errors. Let $\dot{v}_s \in [\dot{v}_{\min}, \dot{v}_{\max}]$ denote the air volume flow rate setpoint for the two fans, where $\dot{v}_{\min}$ and $\dot{v}_{\max}$ is the minimum and maximum achievable air volume flow rate. The cooling coil admits a setpoint $\Delta t$ that represents the reduction of temperature, i.e., $\Delta t = t_o - t_p$, where $t_p$ represents the temperature of the processed air leaving the cooling coil. Let $\Delta t_{\max}$ represent the maximum temperature reduction that can be achieved by the cooling coil. Thus, $\Delta t \in [0, \Delta t_{\max}]$. Let $\alpha \in [0, 1]$ denote the setpoint for the damper system, which is the fraction of the recirculated hot air in the supply air. Thus, $1-\alpha$ is the fraction of the outside air in the supply air. A setpoint $\alpha$ can be achieved by controlling the openness of the three dampers. For example, to achieve $\alpha=0$, the supply and exhaust dampers should be completely open and the mixing damper should be completely closed; to achieve $\alpha=1$, the supply and exhaust dampers should be completely closed and the mixing damper should be completely open.
The control action, denoted by $a$, is a vector $a = [\dot{v}_s, \Delta t, \alpha]$.
We define $A$ as the set of possible actions in the system.

\subsubsection{Reward and costs}
When a control action $a \in A$ is performed at the current time step with a system state of $x \in X$, let $p_{f}(x, a)$ and $p_{c}(x, a)$ denote the average powers consumed by the room fans to maintain the air volume flow rate $\dot{v}_s$ and the cooling coil to lower the temperature by $\Delta t$ Celsius degree over the next control period of $\rho$ seconds, respectively; let $t_{s}(x,a)$ and $\phi_{s}(x, a)$ denote the supply air temperature and RH, respectively.
An immediate reward function, denoted by $r(x, a)$, is defined based on the non-IT power consumption as
\begin{equation}
  r(x, a) = -\big(p_{f}(x, a) + p_{c}(x, a) \big).
  \label{eq:reward}
\end{equation}
Let $t_{\text{th}}$ and $\phi_{\text{th}}$ denote the temperature and RH thresholds for the long-term reliability of the IT hardware equipment, respectively.
To account for the supply air temperature and RH requirement violations as result of taking an action $a$ in a state $x$, we define $c_{t}(x, a)$ and $c_{\phi}(x, a)$ as immediate supply air temperature and RH requirement violation costs. The $c_{t}(x, a)$ and $c_{\phi}(x, a)$ are calculated as
\begin{equation}
\label{eq:cost}
c_{t}(x, a) = \max \left(t_s(x, a) - t_{th}, 0 \right);\quad c_{\phi}(x, a) = \max \left(\phi_s(x, a) - \phi_{th}, 0 \right).
\end{equation}

\subsubsection{Air free-cooled DC control problem}
\label{control_formulation}

At every time step, the system controller observes the system state $x$.
Then, it decides and executes a control action $a$ to operate the supply and exhaust fans, cooling coil and dampers in the next control period of $\rho$ seconds.
At the end of the next control period, the system controller can observe the immediate reward $r(x, a)$ and  supply temperature $c_{t}(x, a)$ and RH $c_{\phi}(x, a)$ costs as feedback signals.
Let $\pi(a|x)$ denote a stationary control policy that maps a state $x \in X$ to an action $a \in A$.
The set of all possible policies is denoted by $\Pi$.
The control design objective is to find a optimal control policy $\pi^*$ that determines $a$ based on $x$ to maximize the expected reward while minimizing temperature and RH requirement violation costs over a long run.
In general, it is difficult to design a closed-form control policy to achieve this goal because the state evolution of the system (i.e., $t_s(x, a)$ and $\phi_s(x, a)$) is complex.

\textbf{MDP formulation:}
The control problem can be formulated as an MDP with an immediate weighted reward function, denoted by $\hat{r}$ as follows:
\begin{equation}
\hat{r}(x, a) = r(x, a) - \zeta_1 c_t(x, a) - \zeta_2 c_{\phi} (x, a).
\label{eq:weighted_reward}
\end{equation}
where $\zeta_1$ and $\zeta_2$ are constant penalty coefficients.
The objective of the above MDP is to find the optimal policy $\pi^*$ that maximizes the long-term expectation of the weighted reward denoted by $\hat{R}(\pi)$. The $\hat{R}(\pi)$ is defined as
\begin{equation}
\hat{R}(\pi) = \mathbb{E}\big[\sum_{\rho=0}^{\infty} \gamma^\rho \hat{r}(x_\rho, a_\rho) \big],
\end{equation}
where $\gamma \in (0, 1)$ is a constant discount factor.
Then, an unconstrained DRL approach can be applied to solve the above MDP problem.
During the interactions between the unconstrained DRL agent and the environment (i.e., the controlled system), the agent is trained to learn the optimal control policy from the historical data including system states, control actions, and the resulted immediate rewards and costs.
With sufficient interactions, the trained agent can well capture the highly complex
system dynamics. Moreover, the learned control policy approaches optimality for a long time horizon
comparable with the time duration of the training phase.
However, due to the usage of the weighted reward $\hat{r}(x, a)$ with constant penalty coefficients,
our evaluation results (cf. \sect\ref{sec:eval}) show
that the unconstrained DRL agent may converge to a control policy which
minimizes the cooling energy or supply temperature/RH requirement violation costs only.

\textbf{CMDP formulation:}
In addition to the above MDP formulation, we also consider a constrained MDP (CMDP) formulation and design a constrained DRL approach to learn the optimal solution of the CMDP.
We define $R(\pi)$, $C_{t}(\pi)$ and $C_{\phi}(\pi)$ as long-term discounted reward,
supply air temperature and RH requirement violation costs, respectively, under a policy $\pi$.
They can be expressed
\begin{align}
R(\pi) = \mathbb{E} \bigg[\sum_{\rho=0}^{\infty} \gamma^\rho r(x_\rho, a_\rho) \bigg],
\quad C_{t}(\pi) = \mathbb{E} \bigg[\sum_{\rho=0}^{\infty} \gamma^\rho c_t(x_\rho, a_\rho)],
\quad C_{\phi}(\pi) = \mathbb{E} \bigg[\sum_{\rho=0}^{\infty} \gamma^\rho c_\phi(x_\rho, a_\rho) \bigg].
\label{eq:expected_reward}
\end{align}
The control design objective now is to find an optimal control policy denoted by $\pi^*$ that maximizes the long-term reward $R(\pi)$ while satisfying the constraints on the long-term costs, i.e., $C_{t} (\pi) \leq 0$ and $C_\phi (\pi) \leq 0$. The CMDP formulation can be formally defined as follows:
\begin{align}
\pi^* = \arg \max_{\pi \in \Pi} R(\pi),
\quad \text{s.t.}\quad C_t(\pi) \leq 0, C_\phi(\pi) \leq 0.
\label{eq:cmdp}
\end{align}
A linear programming (LP)-based approach~\cite{Aitan:1999} can be used to find the optimal policy $\pi^*$ of the above CMDP problem. However, the LP-based approach requires the knowledge of the system state transition probabilities that may not be available due to the complex state evolution of the air free-cooled DC.
Moreover, the model-predictive control (MPC)~\cite{Amos:2018,Amos:2017} can also solve the CMDP optimization problem. However, the MPC is computationally expensive and often for a limited time horizon only.
In this work, to find the optimal control policy of the CMDP defined in Eq.~(\ref{eq:cmdp}),
we apply the Lagrangian relaxation procedure~\cite{Bertsekas:1999} to covert the CMDP problem into an equivalent unconstrained optimization problem as
\begin{equation}
\label{umdp}
\min_{\lambda \geq 0} \max_{\pi_\theta} L(\lambda, \pi_\theta) = \min_{\lambda \geq 0} \max_{\pi_\theta}
\bigg[ R(\pi_\theta) - \lambda_1 C_{t}(\pi_\theta)
- \lambda_2 C_{\phi}(\pi_\theta) \bigg],
\end{equation}
where $\pi_{\theta}$ denotes a control policy characterized by the parameter $\theta$ (e.g., a neural network policy with a collection of weights $\theta$), $L$ is the Lagrangian, $\lambda_i \geq$~(i = 1, 2)
are Lagrangian multipliers (i.e., penalty coefficients) and $\lambda = (\lambda_1, \lambda_2)$.
To solve the unconstrained minmax problem in Eq. (\ref{umdp}),
we design a constrained DRL approach based on the iterative primal-dual policy optimization~\cite{Achiam2017,Liang2018,Chen2018}, in which the primal policy $\pi_\theta$ and dual variables $\lambda$ are iteratively updated in two different timescales. Specifically, on the faster timescale, the policy weight $\theta$ is updated to find $\pi^*_\theta$; on the slower timescale, the dual gradient ascent procedure updates the $\lambda$ until the constraints are satisfied. With sufficient iterations, the primal-dual update procedure can approach the optimal primal-dual solution ($\pi^*_\theta, \lambda^*$).

In \sect\ref{sec:drl}, we will present the detailed design of our DRL system,
including the constrained and unconstrained DRL agents to address the air free-cooled DC control problem.

\section{DRL-based Free-Cooled DC Control}
\label{sec:drl}

\begin{figure}
    \centering
    \includegraphics[width=\linewidth]{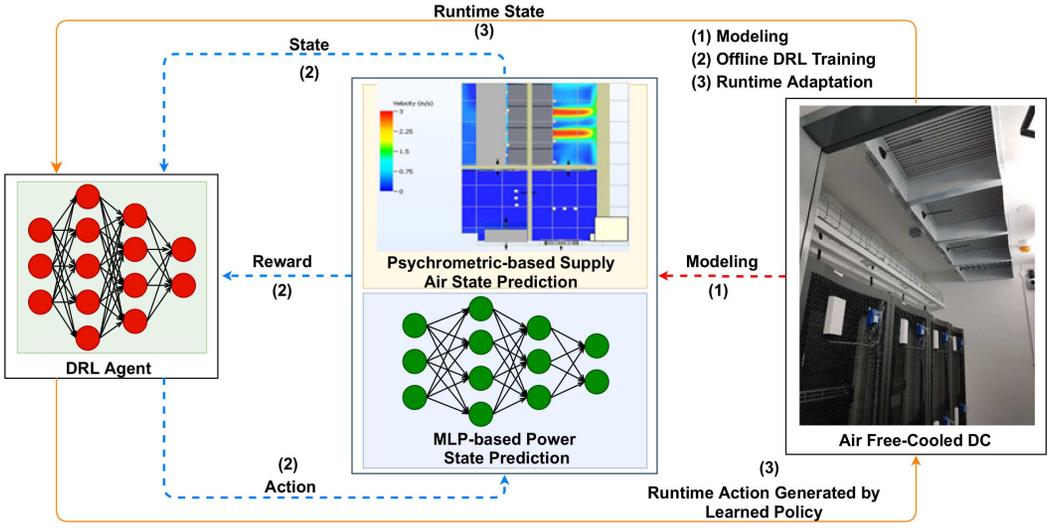}
    \caption{Design workflow of DRL-based DC control.}
    \label{fig:workflow}
\end{figure}

\subsection{Design Workflow}

Typically, DRL agent learns the optimal control policy during the online interactions with the controlled system. However, for free-cooled DC control, the online learning scheme has the following two issues. First, it may take a long time duration to converge, especially when the state and action spaces are large. Second, during the learning phase, excursions due to RL's trial-and-error nature may lead to overheating and server shutdowns.
To address these issues, we adopt an offline DRL training approach in which prediction models are developed to simulate the air free-cooled DC environment.
More specifically, we build psychrometric and two multilayer perceptron (MLP) models to characterize
the state evolution of the supply air temperature and RH. Two other MLPs are developed to model the power consumption of the fans and IT equipment.
In addition, we use the heat energy transform analysis in thermodynamics to model the power of cooling coils.
All these computation models serve as the environment of the DRL agent for offline learning.

Fig.~\ref{fig:workflow} illustrates the workflow of DRL-based DC control approach, which consists of three steps.
First, we collect meta information and real data traces from the air free-cooled DC testbed to train and validate
the prediction models. Second, we use the models validated in the first step to drive the offline training of the DRL agent.
Third, after the completion of the offline training,
the DRL agent is commissioned to control the actual free-cooled DC.

\subsection{Modeling Air Free-Cooled DC}
\label{tdc_modeling}

This section derives the dynamic model that describes the evolution of the steady system state of the air free-cooled DC. We also build two MLPs to characterize the power consumption of servers and supply/exhaust fans.
These models are used for the offline training of DRL agent.

\label{psy_model}

\subsubsection{Dynamic model of supply air temperature and RH state evolution}

In this section, we first perform psychrometric analysis for the four steps of the air processing in the air free-cooled DC as illustrated in Fig.~\ref{fig:top}, i.e., {\em heating} in the server room, {\em buffering} in the buffer chamber, {\em cooling} by the cooling coil,
and {\em mixing} by the damper system.
Based on these models, a Markovian computational model is then constructed to characterize the psychrometric dynamics:
\begin{equation}
\label{eq:psy}
  t_s[k + 1], \phi_s[k + 1] = f(t_s[k], \phi_s[k], t_o[k], \phi_o[k], \dot{v}_s[k], \Delta t, p_{\text{IT}}[k], \alpha)
\end{equation}

We define the following notation: $\dot{m}$ is mass flow rate, $h$ is enthalpy, $w$ is moisture content; for the above psychrometric variables, we use the subscripts $\cdot_s$, $\cdot_h$, $\cdot_{r}$, $\cdot_p$, $\cdot_o$ to refer to the supply air in the cold aisle, the hot air generated by the servers, the recirculated hot air from the buffer chamber to the mixing chamber, the processed air leaving cooling coil, and the outside air provided to the cooling coil, respectively. The four steps are as follows:

{\bf (1) Heating:} Servers generate heat and introduce no extra moisture. Thus, the air enthalpy at the hot aisle is higher than that at the cold aisle, while the moisture contents at the two aisles are identical. Denoting by $\eta$ the servers' heat rate transfer coefficient, the psychrometics of the server room is
\begin{equation}
  \dot{m}_s h_s+\eta p_{\text{IT}} =\dot{m}_h h_h, \quad \dot{m}_s = \dot{m}_h, \quad w_s=w_h.
  \label{eq:server-room}
\end{equation}

{\bf (2) Buffering:} The hot aisle air is transported into the buffer chamber by the exhaust fan. Under the setpoint $\alpha$ for the damper system, the buffer chamber is characterized by
\begin{equation}
  \dot{m}_r = \alpha\dot{m}_h.
  \label{eq:buffer}
\end{equation}

{\bf (3) Cooling:} The total energy of ideal gas is the sum of dry air's energy and water vapor's energy. Without condensation, the cooling coil does not change moisture content of air passing through. Moreover, it does not change mass flow rate. Thus, the condition of the air leaving the cooling coil is given by
\begin{equation}
  h_p=c_p(t_o-\Delta t)+w_p(c_{pw}(t_o-\Delta t)+l), \; w_p = w_o, \; \dot{m}_p=\dot{m}_o,
  \label{eq:cooling}
\end{equation}
where $c_p$ and $c_{pw}$ respectively represent the specific heat of dry air and water vapor which are constants; $l$ represents the evaporation heat. Note that $c_p(t_o-\Delta t)$ is the enthalpy of the dry air leaving the cooling coil; $w_p(c_{pw}(t_o-\Delta t)+l)$ is the enthalpy of the water vapor leaving the cooling coil.

{\bf (4) Mixing:} The air leaving the cooling coil and the recirculated hot air are mixed in the mixing chamber. Governed by the conservation of mass and energy, the psychrometrics of the mixing process can be characterized by
\begin{equation}
  (1-\alpha)h_p+\alpha h_r=h_s, \quad (1-\alpha)w_p+\alpha w_r=w_s, \quad \dot{m}_p+\dot{m}_r=\dot{m}_s.
  \label{eq:mixing}
\end{equation}
Taking the moisture contents of the two influxes as boundaries, Eq.~(\ref{eq:mixing}) suggests that the outflow's moisture content will be in between, which is the basis of the RH control through adjusting $\alpha$.

The above models in Eqs.~(\ref{eq:server-room})-(\ref{eq:mixing}) are for enthalphy, moisture content, and mass flow rate. These quantities can be converted to temperature, RH, and volume flow rate according to the equations presented in~\cite{handbook13}. The aforementioned Markovian computational model is as follows. By initializing the $h_s$ and $w_s$ in Eq.~(\ref{eq:server-room}) with the current state of the supply air condition (i.e., $t_s[k]$ and $\phi_s[k]$), we use the remaining equations in Eqs.~(\ref{eq:buffer})-(\ref{eq:mixing}) to update $h_s$ and $w_s$.
The updated values are then used to initialize the $h_s$ and $w_s$ in Eq.~(\ref{eq:server-room}) again and then solve Eqs.~(\ref{eq:buffer})-(\ref{eq:mixing}). This process is iterated until $h_s$ and $w_s$ converge; the converged values are converted to $t_s[k+1]$ and $\phi_s[k+1]$. Thus, the Markovian computational model has no closed-form expression, presenting a challenge to the design of optimal control policy.

The developed psychrometric analysis-based model can well capture the psychrometric dynamics in the air free-cooled DC and accurately predict the supply air temperature and RH condition.
However, it is computationally expensive due to its internal iterative computation.
Therefore, we also build two MLP models, denoted by MLP$_{t_s}$ and MLP$_{\phi_s}$ which take
$t_s[k]$, $\phi_s[k]$, $t_o[k]$, $\phi_o[k]$, $\dot{v}_s[k]$, $p_{\text{IT}}[k]$, $\Delta t$, and $\alpha$
as inputs to predict $t_s[k + 1]$ and $\phi_s[k + 1]$, respectively.
Our evaluation results in \sect\ref{sec:modelling_accuracy} show that
the MLPs can achieve a similar prediction accuracy as the psychrometric analysis
on real testing data traces collected from the testbed.
However, to achieve high prediction accuracy, the MLPs require a large amount
of training data collected from the real DC.
In addition, due to the RL's trial-and-error nature,
the supply air temperature $t_s$ and RH $\phi_s$ values
under the control policy learned by the DRL agent during the learning phase may
not be within the distribution of the traces of $t_s$ and $\phi_s$ that are used to train the MLPs.
As a result, the MLPs may lead to poor prediction performance during the learning phase of the DRL.
Therefore, the MLP models can be applied for modeling mission-critical DCs
only when we can collect sufficient training data that cover all possible states of the DC.
However, due to the thermal safety requirement of the IT equipment, it can be challenging
to collect such extensive training data.
The psychrometric model can be applied when the coverage of the training data is concerned.

\subsubsection{Power consumption models}
\label{mlp}

We design two MLPs to model the following powers averaged over the next control period: (1) IT power $p_{\text{IT}}[k+1]$, (2) total power of supply and exhaust fans $p_f[k+1]$.
More specifically, to predict the $p_{\text{IT}}[k+1]$, the first power MLP ($\text{MLP}_{\text{IT}}$)
uses the respective IT power measurements in the past $K$ control periods as a part of the input to address the autocorrelation of IT power consumption. This is because the server power has a linear relationship with the server's CPU utilization which is correlated to the server's computing load~\cite{Zhang:2016}.
In this study, with the main focus on the cooling control problem,
we consider the server's CPU utilization as an uncontrollable disturbance.
Therefore, the $\text{MLP}_{IT}$ uses the past IT powers to predict the current IT power.
Moreover, the $\text{MLP}_{IT}$ additionally takes the air volume flow rate $\dot{v}_s[k]$ as inputs
since the air flow generates forces on the blades of the server fans.
The second MLP ($\text{MLP}_f$) modeling $p_f[k+1]$ takes the $\dot{v}[k]$ as inputs.
This is because the fan power increases with the fan speed.
In the literature, the fan power is often modeled as a cubic function of the fan speed.
Note that the hyperparameters of these two MLPs (e.g., the number of layers and neurons)
will be designed in \sect\ref{sec:eval} based on real traces.

We use the heat energy transform analysis in thermodynamics to model the averaged cooling coil power $p_c[k+1]$ over the next control period. Specifically, the $p_c[k+1]$ is modeled as a function of the temperature reduction
$\Delta t[k]$ and air volume flow rate $\dot{v}_s[k]$:
\begin{equation}
\label{aircon_power}
  p_c[k+1] = \frac{c_p \rho \dot{v}_s[k] \Delta t[k]}{\xi},
\end{equation}
whether $\xi \geq 1$ is the thermal coefficient of performance (CoP) of the cooling coil, $c_p = 1.006$ kJ/kg\textdegree{}C is the specific heat of air and $\rho = 1.202$ kg/$\text{m}^3$ is the density of air.
Eq.~(\ref{aircon_power}) models the electrical power $p_c[k+1]$ that the cooling coil consumes
to lower the temperature of the outside air by $\Delta t$\textdegree{}C at the air flow rate $\dot{v}_s[k]$.
The cooling coil consumes more power when (1) a higher cooling capacity $\Delta t$ is required
and (2) a larger volume of the outside air is processed.
The CoP $\xi$ characterizes how efficient the cooling coil uses its electrical power input
to remove the heat from its processing air.

The average non-IT power consumed in the next control period is $p[k+1] = p_f[k+1] + p_c[k+1]$.

\begin{table}[b]
\renewcommand{\arraystretch}{1.2}
\caption{List of notation used in DRL agent training}
\label{drl_notation}
\centering
\begin{tabular}{|l|l|}
\hline\
\textbf{Notation} & \textbf{Definition}\\
\hline
$Q(x, a; \theta)$,  $Q'(x, a; \theta')$ & Primary and target DQNs\\
\hline
$\theta[k]$, $\theta'[k]$ & Weight collection of primary and target DQNs at time step $k$\\
\hline
$x[k]$, $a[k]$, $r[k]$ & State, action and reward at the $k$th time step\\
\hline
$t_s[k]$, $\phi_s[k]$ & Supply air temperature and RH at time step $k$\\
\hline
$p_{\text{IT}}[k]$, $p_f[k]$, $p_c[k]$ & IT, fan and cooling coil powers at time step $k$\\
\hline
$\hat{r}[k]$, $c_t[k]$, $c_{\phi}[k]$ & Weighted reward, temperature and RH costs at time step $k$\\
\hline
$\zeta_1$, $\zeta_2$ & Constant penalty coefficients of unconstrained DRL agent\\
\hline
$\lambda_{1}[k]$, $\lambda_{2}[k]$ & Penalty coefficients of constrained DRL agent at time step $k$\\
\hline
$\eta_1[k]$, $\eta_1[k]$ & Learning step sizes at time step $k$\\
\hline
$\bar{t_s}[k]$, $\bar{\phi_s}[k]$ & Averaged supply air temperature and RH at time step $k$\\
\hline
$\gamma$, $\beta$ & Discount factor and parameter of soft target DQN update\\
\hline
\end{tabular}
\end{table}

\subsection{Design of Offline DRL Agents}
In this section, we adopt the training framework~\cite{DQN2015}
to train offline DQNs for the unconstrained and constrained DRL agents
to capture good control policies with objectives defined in~\sect\ref{control_formulation}.
Definitions of main notation used in the training of DRL agents are presented in Table~\ref{drl_notation}.

\subsubsection{Unconstrained DRL agent}
To learn the optimal policy of the MDP problem, the DQN denoted by $Q(x, a; \theta)$,
and its target network denoted by $Q'(x, a; \theta')$ are
iteratively updated by interacting with DC environment that
is modeled by the computational models developed in~\sect\ref{tdc_modeling}.
The detailed training procedure of the unconstrained DRL agent is presented in Algorithm~\ref{alg:uDRL_training}.
The training phase lasts for $N$ episodes, each of which consists of $T$ time steps.
An episode starts with a state chosen randomly from the training data.
At the beginning of the $k$th time step, an action $a[k]$ is selected for state $x[k]$ according to the $\varepsilon$-greedy algorithm~\cite{Sutton1998}
based on action-values $Q(x, a)$, given by the current $Q(x, a; \theta)$.
The immediate weighted reward $\hat{r}[k]$ is obtained as a feedback signal from the modeled DC environment. Specifically, given the selected action $a[k]$, the $t_s[k+1]$, $\phi_s[k+1]$ and $p_{\text{IT}}[k+1]$ are estimated using the psychrometric model and IT power model (i.e., MLP$_\text{IT}$),
where the outside air condition (i.e., $t_o[k+1]$ and $\phi_o[k+1]$) are taken from real sensor measurements.
The powers of fans $p_f[k+1]$ and cooling coil $p_c[k+1]$ with respect to the selected $\dot{v}_s[k]$ and $\Delta t[k]$ are determined using MLP$_\text{f}$ and the thermodynamic-based cooling coil power model, respectively.
The costs $c_t[k]$ and $c_{\phi}[k]$ are calculated based on the $t_s[k+1]$ and $\phi_s[k+1]$ using Eqs.~(\ref{eq:cost}), respectively.
The $\hat{r}[k]$ is calculated based on $p_f[k+1]$, $p_c[k+1]$, $c_t[k]$ and $c_{\phi}[k]$ with constant values
of $\zeta_1$ and $\zeta_2$.

\begin{algorithm}[t]
\caption{Offline training of unconstrained DRL agent.}
\label{alg:uDRL_training}
Initialize DQN-network $Q(x, a; \theta)$ and its target network $Q'(x, a; \theta')$ with random weights\\
Initialize empty replay memory $\Omega$\\
    \For{n = 1, 2, \ldots, N}{
        Obtain initial state $x[1]$ from real data traces\\
        \For{k = 1, 2, \ldots, T}{
            Select action $a[k]$ for state $x[k]$ according to the $\epsilon$-greedy algorithm\\
            Execute action $a_k$ and observer $\hat{r}[k]$, and $x[k+1]$\\
            Store transition $\big( x[k], a[k], \hat{r[k]}, x[k+1] \big)$ to the memory $\Omega$\\
            Sample a random minibatch of $\big\{(x[i], a[i], \hat{r}[i], x[i+1]) \big\}_{i=1}^M$ from the $\Omega$\\
            Set $y[i] = \hat{r}[i] + \gamma \, \underset{a}{\max}\, Q'(x[i+1], a; \theta')$\\
            Update the $Q(s, a; \theta[k])$ with loss $L_\theta[k] = \frac{1}{M} \sum_{i=1}^{M}\left( y[i] - Q(x[i], a[i]; \theta[k]) \right)^2$ by performing the gradient descent as\\
            $\theta[k+1] = \theta[k] - \eta_1 \nabla L_{\theta[k]}$\\
            Update the target network $Q(x, a; \theta'[k])$ by setting $\theta'[k+1] = \beta\theta[k] + (1-\beta)\theta'[k]$\\
        }
    }
\end{algorithm}

The experience replay mechanism is used to update the DQN during the training
as presented in lines 8-12 of Algorithm~\ref{alg:uDRL_training}.
Let $\theta[k]$ and $\theta'[k]$ denote the collection of weights of the DQN and its target network
at the $k$th time step. At every time step $k$, the transition tuple $\big(x[k], a[k], \hat{r}[k], x[k+1] \big)$ is stored in the experience replay memory with the size of $\Omega$.
To update the $Q(x, a; \theta[k+1])$ which is used to determine state-action values $Q(x, a)$
in next time step, a random mini-batch of $M$ transitions $\big\{ (x[i], a[i], \hat{r}[i], x[i+1])\big\}_{i=1}^M$ is sampled from the replay memory.
Then, the weights $\theta[k+1]$ is updated by performing a gradient descent step using $M$ mini-batch samples
as follows:
\begin{equation}
\label{eq:policy_update}
\theta[k+1] = \theta[k] - \eta_1 \nabla L_{\theta[k]},
\end{equation}
where $\eta_1 \leq 0$ is the step size of the gradient descent and $\nabla L_{\theta[k]}$ is the gradient of the loss function denoted by $L_\theta[k]$. The $L_{\theta[k]}$ is defined as
\begin{equation}
\label{eq:loss}
L_\theta[k] = \frac{1}{M} \sum_{i = 1}^M \bigg(y[i] - Q(x[i], a[i]; \theta[k]) \bigg)^2,
\end{equation}
where the target $y[i] = \hat{r}[i] + \gamma \, \underset{a}{\max} \, Q'(x[i+1], a; \theta'[k])$
is calculated based on the weighted reward $\hat{r}[i]$ and the target network $Q'(x, a; \theta'[k])$.
Finally, we use the soft target update method~\cite{Timothy:2016} to update the weights of the target Q-network by setting $\theta'[k+1] = \beta \theta[k+1] + (1-\beta)\theta'[k]$ with $0 < \beta \ll 1$.
The soft target update often gives better learning stability than the hard target update of the original DQN training.

\begin{algorithm}[t]
\caption{Offline training of constrained DRL agent.}
\label{alg:cDRL_learning}
Initialize DQN-network $Q(x, a; \theta)$ and its target network $Q'(x, a; \theta')$ with random weights\\
Initialize empty replay memory $\Omega$ and set $\lambda_1 = \lambda_2 = 0$\\
    \For{n = 1, 2, \ldots, N}{
        Obtain initial state $x[1]$ from real data traces\\
        \For{k = 1, 2, \ldots, T}{
            Select action $a[k]$ for state $x[k]$ according to the $\epsilon$-greedy algorithm\\
            Execute action $a_k$ and observer $r[k], c_t[k], c_\phi[k]$ and $x[k+1]$\\
            Store transition $\big( x[k], a[k], r[k], c_t[k], c_\phi[k], x[k+1] \big)$ to the $\Omega$\\
            Sample a random minibatch of $\big\{(x[i], a[i], r[i], c_t[i], c_\phi[i], x[i+1]) \big\}_{i=1}^M$ from the $\Omega$\\
            Set $y[i] = r[i] - \lambda_1[k] c_t[i] - \lambda_2[k] c_\phi[i] + \gamma \, \underset{a}{\max} \, Q'(x[i+1], a; \theta')$\\
            Update the $Q(s, a; \theta[k])$ with loss $L_\theta[k] = \frac{1}{M} \sum_{i=1}^{M}\left( y[i] - Q(x[i], a[i]; \theta[k]) \right)^2$ by performing the gradient descent as\\
            $\theta[k+1] = \theta[k] - \eta_1 \nabla L_{\theta[k]}$\\
            Update penalty coefficients by setting:\\
            $\lambda_{1}[k + 1] = \Gamma_{\lambda_1} \big[\lambda_{1}[k] + \eta_2 \big(\bar{t}[k] - t_{th}) \big]$\\
            $\lambda_{2}[k + 1] = \Gamma_{\lambda_2} \big[\lambda_{2}[k] + \eta_2 \big(\bar{\phi}[k] - \phi_{th})\big]$\\
            Update the target network $Q(x, a; \theta'[k])$ by setting $\theta'[k+1] = \beta\theta[k] + (1-\beta)\theta'[k]$\\
        }
    }
\end{algorithm}

\subsubsection{Constrained DRL agent}

The constrained DRL agent is trained based on the primal-dual update method in which the weight $\theta$
and penalty coefficient $\lambda_1$  and $\lambda_2$ are iteratively updated at every time step.
The primal-dual update procedure for the constrained DRL agent also uses the learning framework of the DQN.
Algorithm~\ref{alg:cDRL_learning} presents the update procedure.
Different from the unconstrained DRL approach that uses the constant penalty coefficients, the constrained DRL approach updates the penalty coefficients $\lambda_1$ and $\lambda_2$ during the training until the requirements on the supply air temperature and RH are satisfied. Moreover, at every training step, the weights of the DQN networks are updated using the last updated values of $\lambda_1$ and $\lambda_2$.

At every time step $k$, the reward $r[k]$ and costs $c_t[k]$ and $c_\phi[k]$ are stored in the reply memory.
Denote by $\lambda_1[k]$ and $\lambda_2[k]$ the supply air temperature and RH penalty coefficients respectively,
at the $k$th time step. Then, the weight $\theta[k+1]$ of the $Q(x, a; \theta[k])$ in next time step is updated
by minimizing the loss function $L_{\theta}[k]$ with the targets $y[i]$ ($i = 1,\ldots, M$)
are calculated as follows:
\begin{equation}
\label{eq:target}
y[i] = (r[i] - \lambda_1 c_t[k]) - c_\phi[k] + \gamma \, \underset{a}{\max} \, Q'(x[i+1], a; \theta'[k]).
\end{equation}
The detailed procedure for updating the $\theta[k+1]$ is presented in lines 9-12 of Algorithm~\ref{alg:cDRL_learning}.
Let $\bar{t_s}[k] = \frac{1}{W}\sum_{i = k-W}^{k} t_s[i]$ and $\bar{\phi_s}[k] = \frac{1}{W} \sum_{i=k-W}^{k} \phi_s[i]$ denote the average of supply air temperature and RH over a horizon window of $W$.
At every time step $k$, the $\lambda_1$ and $\lambda_2$ are updated with a step size $\eta_2[k] \leq 0$ as
\begin{align}
\lambda_{1}[k + 1] &= \Gamma_{\lambda_1} \big[\lambda_{1}[k] + \eta_2 \big(\bar{t}[k] - t_{th}) \big],
\label{eq_coefficient1}\\
\lambda_{2}[k + 1] &= \Gamma_{\lambda_2} \big[\lambda_{2}[k] + \eta_2 \big(\bar{\phi}[k] - \phi_{th}) \big],
\label{eq_coefficient2}
\end{align}
where $\Gamma_{\lambda_i}$ (i = 1, 2) represent projection operators that keep $\lambda_1$ and $\lambda_2$ within intervals of $[0, \lambda_i^{max}]$ for large positive constants $\lambda_i^{max} < \infty$.
During the learning phase, we set $\eta_1[k] > \eta_2[k]$ to enable that the policy weight $\theta$ is updated on a faster timescale than that of the coefficients $\lambda_1$ and $\lambda_2$.
To guarantee the convergence of primal-dual policy update to the optimal solution ($\pi^*_\theta, \lambda^*$),
the step sizes $\eta_1$ and $\eta_2$ are required to satisfy the standard conditions
for stochastic approximation algorithms~\cite{chow:2017} as follows:
\begin{equation}
\label{eq_step_size_condition}
\sum_{k=1}^{\infty} \eta_1[k] =  \sum_{k=1}^{\infty} \eta_2[k] = \infty,
\quad \sum_{k=1}^{\infty} \eta_1[k]^2 + \eta_2[k]^2 < \infty,
\quad \lim_{k \rightarrow \infty} \frac{\eta_2[k]}{\eta_1[k]} = 0,
\end{equation}
where $\eta_1[k]$ and $\eta_2[k]$ are values of step sizes at the $k$th time step.
In this study, we keep $\eta_1$ and $\eta_2$ at constant values during the training phase.
The constrained DRL approach also uses the soft target update method
to update the weights of the target Q-network at every time step $k$.

\subsection{Sensor and Experiment Requirements}
\label{subsec:requirement}

The testbed presented in \sect\ref{subsec:testbed} is instrumented with many sensors to monitor the system state. To run trained DRL agents, the essential sensors include: (1) temperature and RH sensors to monitor the outside air and supply air conditions; (2) a meter to monitor the total power consumption of the IT equipment. Moreover, to implement the primary controls of the supply/exhaust fans, the cooling coil, and the damper system, we need the following sensors: (1) air volume flow rate sensors to monitor the air entering the cold aisle and the air passing the mixing damper;
(2) a temperature sensor measuring the air leaving the cooling coil.
To collect training data for the offline learning of DRL, meters to measure the power consumption of supply and exhaust fans, as well as the cooling coil are needed in addition to the sensors mentioned above.
More specifically, the air free-cooled DC operators who adopt our DRL control approaches need to conduct controlled experiments for
collecting the measurements of the supply air temperature $t_s$ and RH $\phi_s$, IT power $p_{\text{IT}}$, fan power $p_{\text{f}}$ and cooling coil power $p_{c}$.
In these experiments, the input parameters including air volume flow rate $\dot{v}_s$, cooling capability $\Delta t$,
and air mixing ratio $\alpha$ are controlled in wide ranges under real outside air conditions,
such that the collected measurement data can cover all possible DC environment running conditions
under the air free cooling setting in the tropics.
Note that conducting such controlled experiments is a one-time effort in the DRL learning phase only.
Moreover, the tropical air free-cooled DC scheme caters into enterprise DCs who own the IT equipment.
As such, they have full control of the DC as well as the IT equipment to perform these required experiments to apply our approach.

\section{Performance Evaluation}
\label{sec:eval}

This section evaluates the system state prediction and the DRL-based controller using simulations driven by real data traces collected from the air free-cooled testbed. The psychrometric model is implemented using Matlab 2019.
The MLP models and DRL agents are
implemented in Python 3.5 with Keras 2.1.6 using TensorFlow 1.8.0.
Specifically, we adopt the implementation framework of the Keras-RL project~\cite{keras-rl}
to implement the DQN training and optimization process for our DRL agents.

\subsection{Accuracy of Air Free-Cooled DC Modeling}
\label{sec:modelling_accuracy}

\begin{figure}[t]
\centering
\subfloat[Temperature prediction by the psychrometric model and MLP$_{t_s}$ over 100 samples.
Their prediction RMSEs over 1059 samples are 0.76\textdegree{}C and 0.57\textdegree{}C, respectively.]
{\includegraphics[width=\columnwidth]{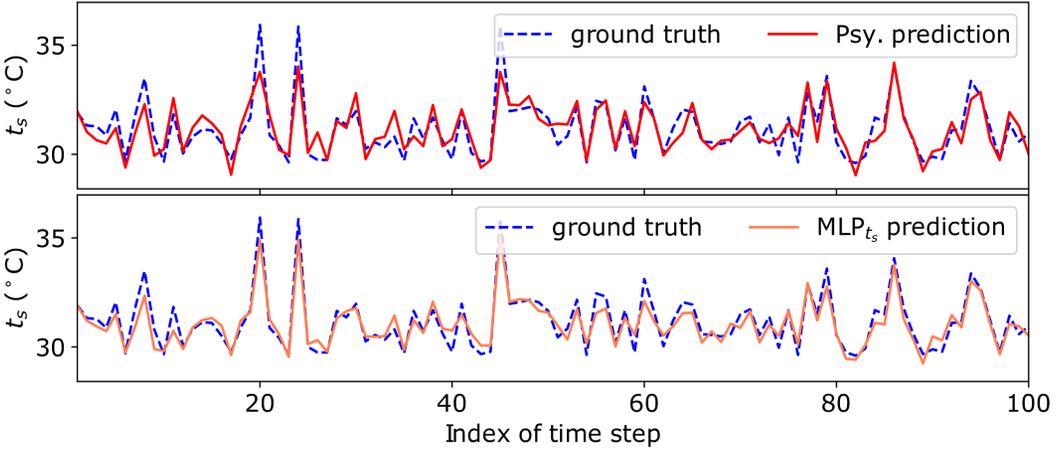}}\\
\quad\\
\subfloat[RH prediction by the psychrometric model and MLP$_{\phi_s}$ over 100 samples.
Their prediction RMSEs over 1059 samples are 6.9\% and 5.4\%, respectively.]{\includegraphics[width=\columnwidth]{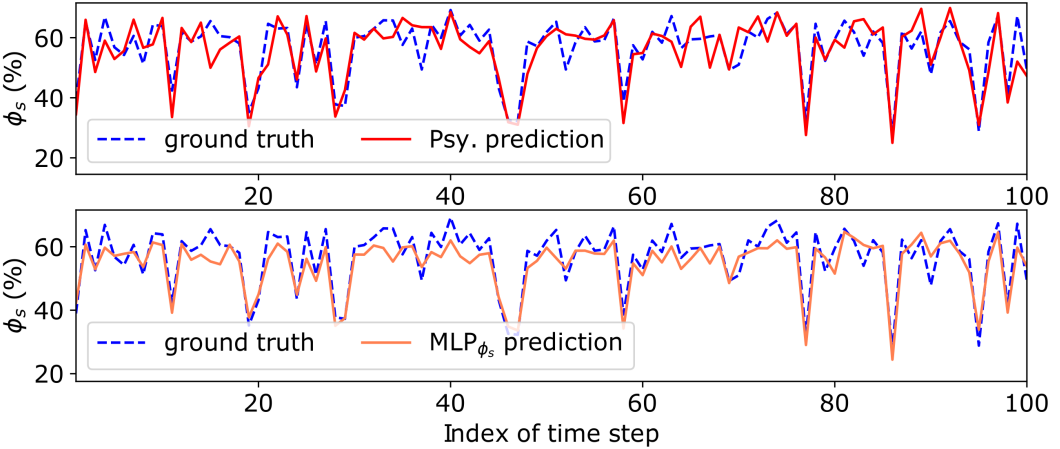}}
\caption{Prediction performance of the psychrometric model and MLPs.}
\label{fig:psy_mixing}
\end{figure}

\subsubsection{Model of supply air temperature and RH state evolution}
We use data traces collected during the controlled experiments
on the testbed (cf.~\ref{subsubsec:impact-performance}) to evaluate the psychrometric model and two MLPs (i.e., MLP$_{t_s}$ and MLP$_{\phi_s}$) presented in \sect\ref{psy_model}.
The inputs to these models are $t_o$, $\phi_o$, $p_{\text{IT}}$, $\dot{v}_s$, $\Delta t$, and $\alpha$; the outputs are the predicted $t_s$ and $\phi_s$.
We use root mean squared error (RMSE) between the prediction and the ground truth as the performance metric
to evaluate the psychrometric model and two MLPs.
More specifically, the MLP$_{t_s}$ and MLP$_{\phi_s}$
are trained and validated using 2,120 and 1,060 data samples, respectively.
We perform various evaluation to select the optimal numbers of hidden layers and neurons
to minimize the prediction RMSEs on the testing data.
The MLP$_{t_s}$ and MLP$_{\phi_s}$ with 20 hidden layers, each of which consists of 5 neurons,
are chosen since they achieve the smallest testing RMSEs among evaluated MLPs.
Fig.~\ref{fig:psy_mixing} shows the prediction results over a time duration of 1.67 hours (i.e., 100 samples).
We can see that the prediction by the psychrometric model and two MLPs well tracks the ground truth.
The RMSEs of the psychrometric model for $t_s$ and $\phi_s$ are 0.76\textdegree{}C and 6.9\%,
respectively, over an evaluated period of 17.65 hours (i.e., 1059 samples).
The MLP$_{t_s}$ and MLP$_{\phi_s}$ have slightly smaller RMSEs of 0.57\textdegree{}C and 5.4\%
for $t_s$ and $\phi_s$, respectively, than those of the psychrometric model.
However, note that the MLPs require to be trained on a large number of training and validating data samples,
while the psychrometric model can achieve a good prediction performance without training.
Moreover, the MLPs may have a poor prediction performance on testing data
which are not within the distribution of the training data.
Compared with the psychrometric model, the MLPs can be a better choice
only when a sufficient number of training samples can be collected.
Due to unforseen thermal conditions of air free-cooled DCs caused by the RL's trial-and-error nature during the learning phase, we will use the psychrometric model to train and evaluate our unconstrained and constrained DRL agents in \sect\ref{sec:DRL_training}.

\begin{figure}[t]
\centering
\includegraphics[width=\columnwidth]{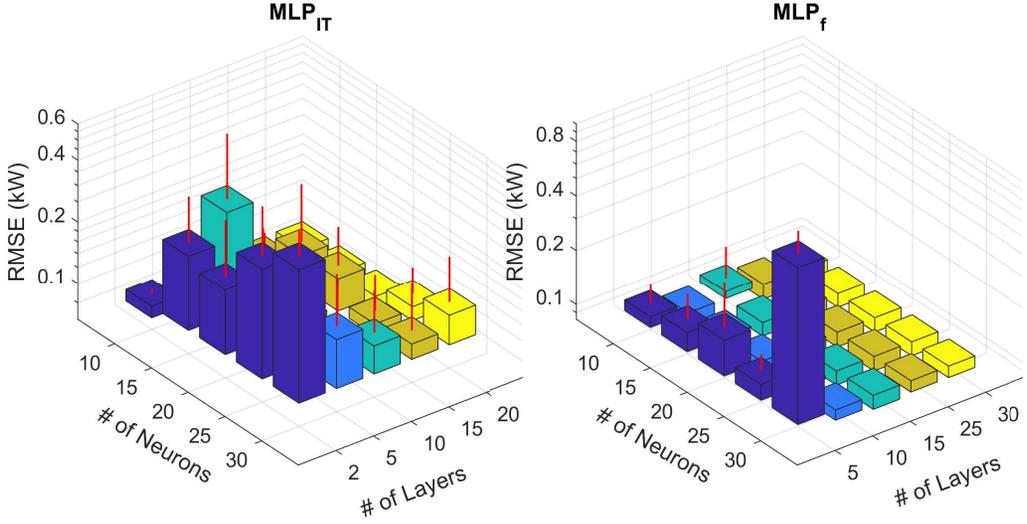}
  \caption{Impact of hyperparameters on MLPs' performance (z-axis is in log scale).}
  \label{fig:mlp_rmse}
\end{figure}

\begin{figure}
  \centering
  \includegraphics[width=\columnwidth]{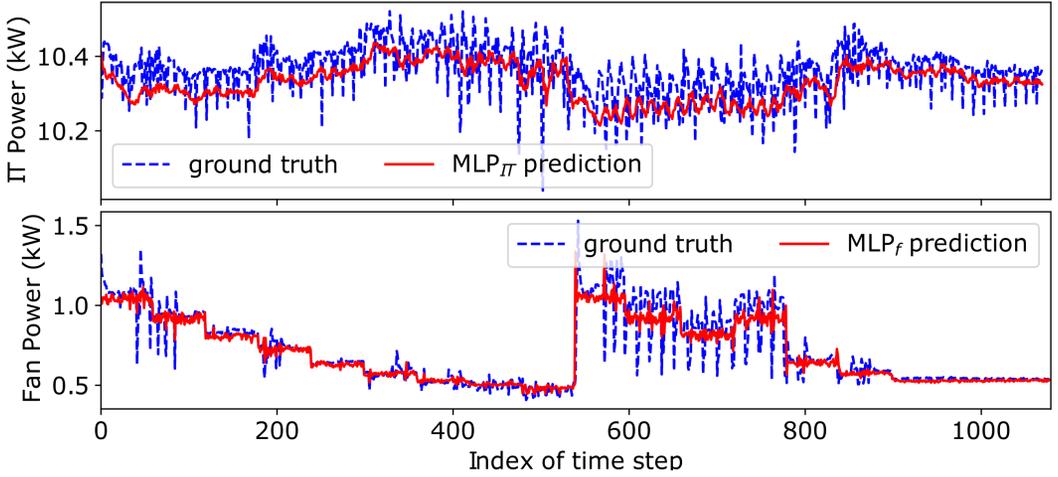}
  \caption{Prediction results of MLPs. Top: IT power; bottom: fan power.}
  \label{fig:mlp_prediction}
\end{figure}

\subsubsection{MLP-based power prediction}

We evaluate the two MLP models presented in \sect\ref{mlp} for predicting IT power and fan power.
Each MLP is trained, validated and tested using 1375, 700 and 1080 data samples, respectively.
The settings of $K$ for the MLP$_{\text{IT}}$ is 10. That is, the power measurements in the immediate past 10 control periods are used for predicting the IT power in the next control period.
For all MLPs, the training batch size is set to 128;
the training time is 1,000 epochs. The Adam optimizer with a learning rate of 0.001 is used for training. Moreover, we use the rectified linear units (ReLUs) as the activation function for input and hidden layers; we use linear units for output layer. We conduct extensive evaluation to choose the number of hidden layers and neurons for each MLP to minimize the prediction RMSEs.
The evaluation for a certain combination of hyperparameter settings is repeated 5 times to account for the randomness of the training.
Fig.~\ref{fig:mlp_rmse} shows the error bars for testing RMSEs with various hyperparameter settings of the number of hidden layers and the number of neurons. $\text{MLP}_{IT}$ achieves the smallest RMSE of $0.66 \pm 0.01\,\text{kW}$ with 5 hidden layers, each of which has 10 neurons.
$\text{MLP}_f$ achieves the smallest RMSE of $0.08 \pm 0.006\,\text{kW}$ with 25 layers, each of which has 10 neurons. Fig.~\ref{fig:mlp_prediction} shows the ground truth and the prediction by the $\text{MLP}_{IT}$ and $\text{MLP}_f$ with the chosen hyperameters over a time duration of 18 hours. Overall, the predictions well track the ground truths.

\subsection{DRL Agent Training and Execution}
\label{sec:DRL_training}

\subsubsection{Settings}

We build unconstrained and constrained DQNs (i.e., the primary and target action-value functions) as fully-connected deep neural networks with the same network architecture. Each network consists of an input layer, three hidden layers and a linear output layer. The first hidden layer has 128 ReLUs while the second and third layers have 64 and 32 ReLUs, respectively. From our extensive trials, the choice of three-layer perception with selected (128, 64, 32) ReLU neurons achieves satisfactory convergence performance for unconstrained and constrained DRL-based approaches for the simulated air free-cooled DC.
Both unconstrained and constrained DRL agents admit a system state and choose an action $a = [\dot{v}_s, \Delta t, \alpha]$ from a discrete action space: $\dot{v}_s$ is from 2000~$\text{m}^3/\text{h}$ to 10000~$\text{m}^3/\text{h}$ with step size of 2000~$\text{m}^3/\text{h}$; $\Delta t$ is from 0\textdegree{}C to 15\textdegree{}C with step size of 1\textdegree{}C; and $\alpha$ is from 0 to 1 with step size of 0.1.
These step sizes are from the physical constraints of the supply/exhaust fans,
the cooling coil, and the damper system. The size of the action space is $5 \times 16 \times 11 = 880$.
For the cooling coil power model, we set the CoP $\xi$ to 2.
The control period is one minute. For the offline training of constrained and unconstrained DQNs,
we adopt the following settings: training batch size is 64; replay memory size is 50000;
sort target update weight $\beta = 0.01$;
the $\varepsilon$ of the $\varepsilon$-greedy method reduces linearly from 1 to 0.1.
Moreover, the steps $\eta_1$ and $\eta_2$ are fixed at 0.01 and 0.001, respectively.
We set the discount factor $\gamma$ for the unconstrained and constrained DRL agents to 0.99 and 0.5, respectively.
For the unconstrained DRL approach, the penalty coefficients $\zeta_1$ and $\zeta_2$ are set to 2.
For the constrained DRL approach, the window size $w$ is set to 50 and
the $\lambda_1^{\max}$ and $\lambda_2^{\max}$ are set to 100.

\begin{table}[t]
\renewcommand{\arraystretch}{1.3}
\caption{Parameters and Values}
\label{sim_parameter}
\centering
\begin{tabular}{|l|l|l|}
\hline\
\textbf{Parameters} & \textbf{Values}\\
\hline
Air flow rate $\dot{v}$ & [2000, 10000] $\text{m}^3/\text{h}$\\
\hline
Temperature reduction $\Delta t$ & [0\textdegree{}C, 15\textdegree{}C]\\
\hline
Mixing damper system setpoint $\alpha$ & [0, 1]\\
\hline
Cooling coil CoP $\xi$ & 2\\
\hline
Temperature threshold $t_{th}$ & [32\textdegree{}C, 35\textdegree{}C, 40\textdegree{}C] \\
\hline
RH threshold $\phi_{th}$ & [65\%, 80\%] \\
\hline
Decision period $\rho$ & 1 min\\
\hline
Exploration $\epsilon$ & [1, 0.1]\\
\hline
Replay memory and minibatch sizes ($\Omega, M$) & (50000, 64)\\
\hline
Learning step sizes ($\eta_1, \eta_2$)& (0.01, 0.001)\\
\hline
Unconstrained DRL's penalty coefficients $(\zeta_1, \zeta_2)$ & (2, 2)\\
\hline
Maximum penalty coefficients ($\lambda_1^{\max}$ and $\lambda_2^{\max}$) & (100, 100)\\
\hline
Number of training episodes and steps ($N, T$) & (3000, 1000)\\
\hline
\end{tabular}
\end{table}

We evaluate the performance of two DRL approaches under wide ranges of the settings for
supply air temperature and RH requirements. Specifically, we vary the temperature threshold $t_{\text{th}}$ in a range consisting of 32\textdegree{}C, 35\textdegree{}C and 40\textdegree{}C, which are the reliability temperatures of ASHRAE Class A1, A2, and A3 servers, respectively. The RH threshold $\phi_{\text{th}}$ is varied from 65\% to 80\%. Note that the 65\% is the deliquescent RH of many contaminants \cite{ASHRAE_Contamination},
while the 80\% is the reliability RH of ASHRAE Class A1 server.
Major simulation parameters are summarized in Table~{\ref{sim_parameter}.

\begin{figure}
  \centering
  \includegraphics[width=\columnwidth]{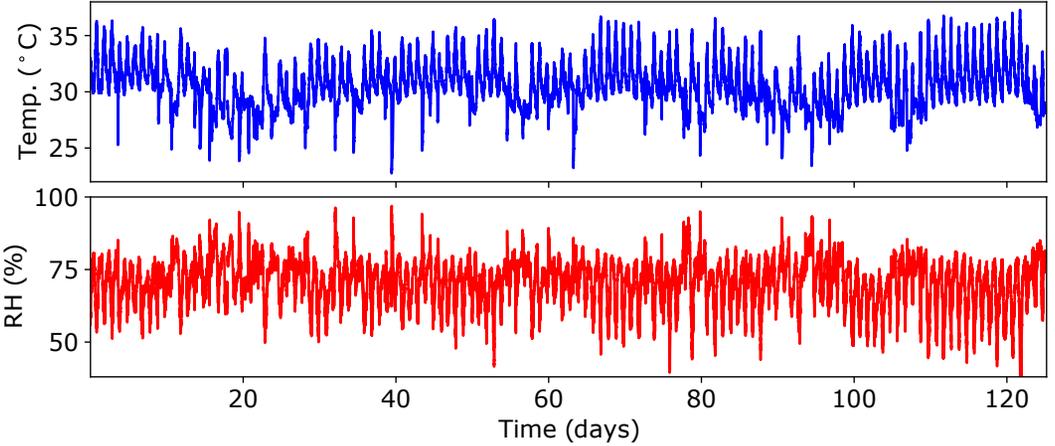}
  \caption{Outside temperature and RH condition of our testbed area over 125 days.}
  \label{fig:outside_condition}
\end{figure}

\subsubsection{DRL agent training}
In this section, we evaluate the learning performance of the proposed unconstrained and constrained DRL approaches under various settings for $t_{th}$ and $\phi_{th}$.
In this section, we use uDRL and cDRL to denote the two approaches.
Fig.~\ref{fig:outside_condition} shows 125 days' outdoor air conditions of our testbed area. We use the first 118 days' data for training the DRL agents and the remaining data for evaluating the trained agents.
The offline training is for $N = 3,000$ episodes, each of which consists of $T = 1,000$ control periods.
At the beginning of each episode, we select a batch of 1,000 samples of outside air condition to drive the training. During the training, the system state is determined based on the action taken by the agent, the psychrometric model, and the power models presented in \sect\ref{sec:drl}.

\begin{figure}
  \centering
  \begin{subfigure}{.325\textwidth}
    \includegraphics[width=\textwidth]{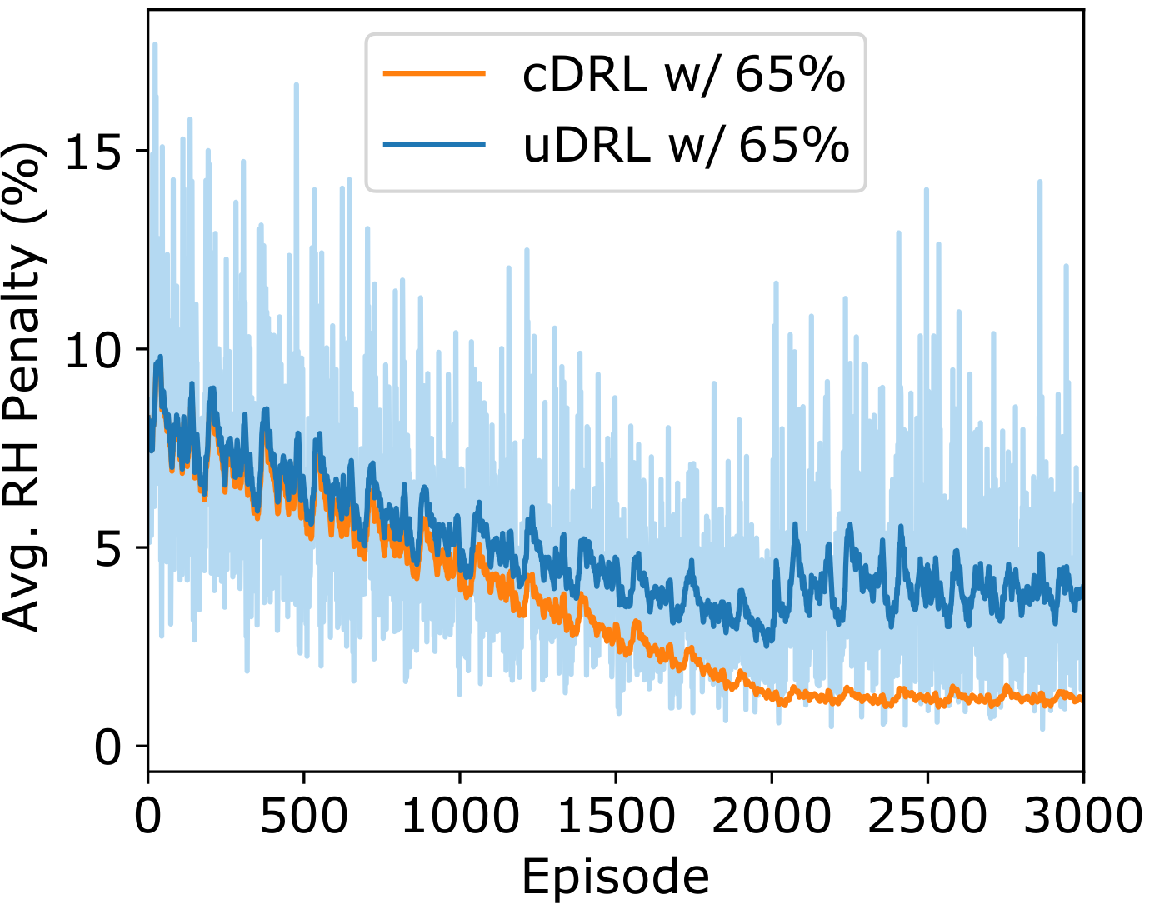}
    \caption{RH penalty.}
    \label{fig:training_rh_65}
  \end{subfigure}
  \begin{subfigure}{.325\textwidth}
    \includegraphics[width=\textwidth]{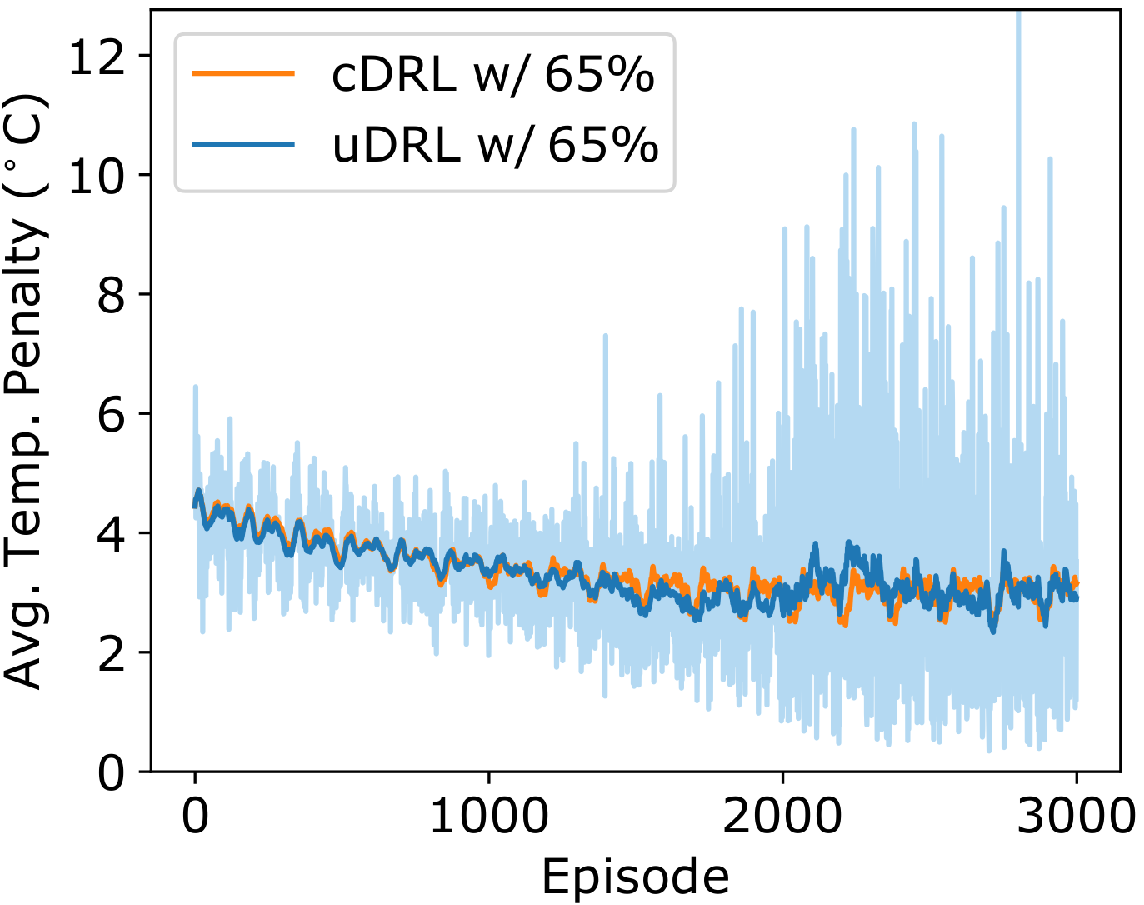}
    \caption{Temperature penalty.}
    \label{fig:training_temp_65}
  \end{subfigure}
  \begin{subfigure}{.325\textwidth}
    \includegraphics[width=\textwidth]{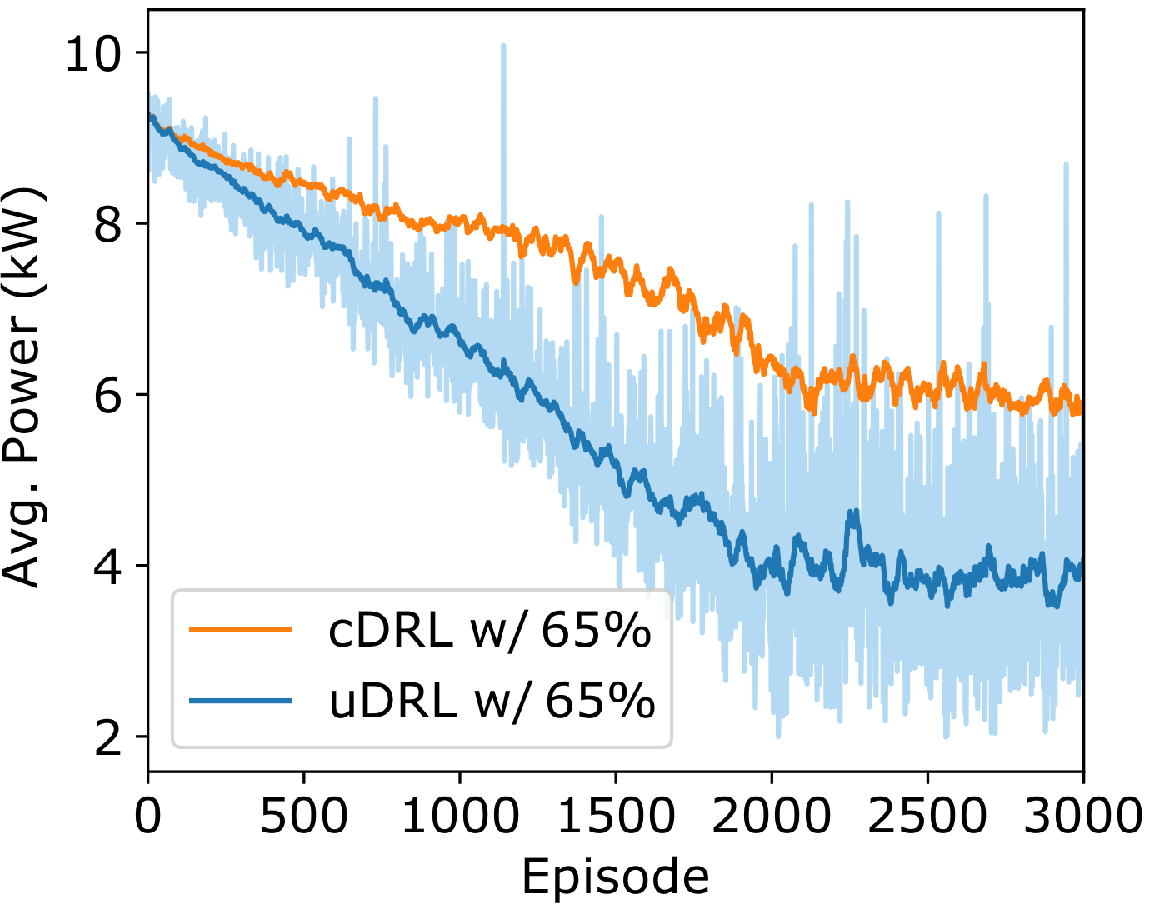}
    \caption{Cooling power.}
    \label{fig:training_energy_65}
  \end{subfigure}
  \caption{Training performance with $t_{\text{th}}$ = 32\textdegree{}C and $\phi_{\text{th}} = 65\%$.}
  \label{fig:training_65}
\end{figure}

Figs.~\ref{fig:training_65} and \ref{fig:training_80} show training traces, including average RH and temperature penalties
(i.e., $\max(t_s(x, a) - t_{th}, 0)$ and $\max(\phi(x, a) - \phi_{th}, 0)$) and average cooling powers of the cDRL and uDRL approaches over an episode of 1,000 time steps when the $t_{th}$ is 32\textdegree{}C and the $\phi_{th}$ are 65\% and 80\%. Along the training episodes, with both uDRL and cDRL  approaches, the power consumption and RH and temperature penalties have increasing variance but decreasing overall trend.
They mostly drop during training. The results show that the training of both cDRL and uDRL agents tends to be convergent after a certain number of training episodes (e.g., $N = 3,000$).

\begin{figure}
  \centering
  \begin{subfigure}{.32\textwidth}
    \includegraphics[width=\textwidth]{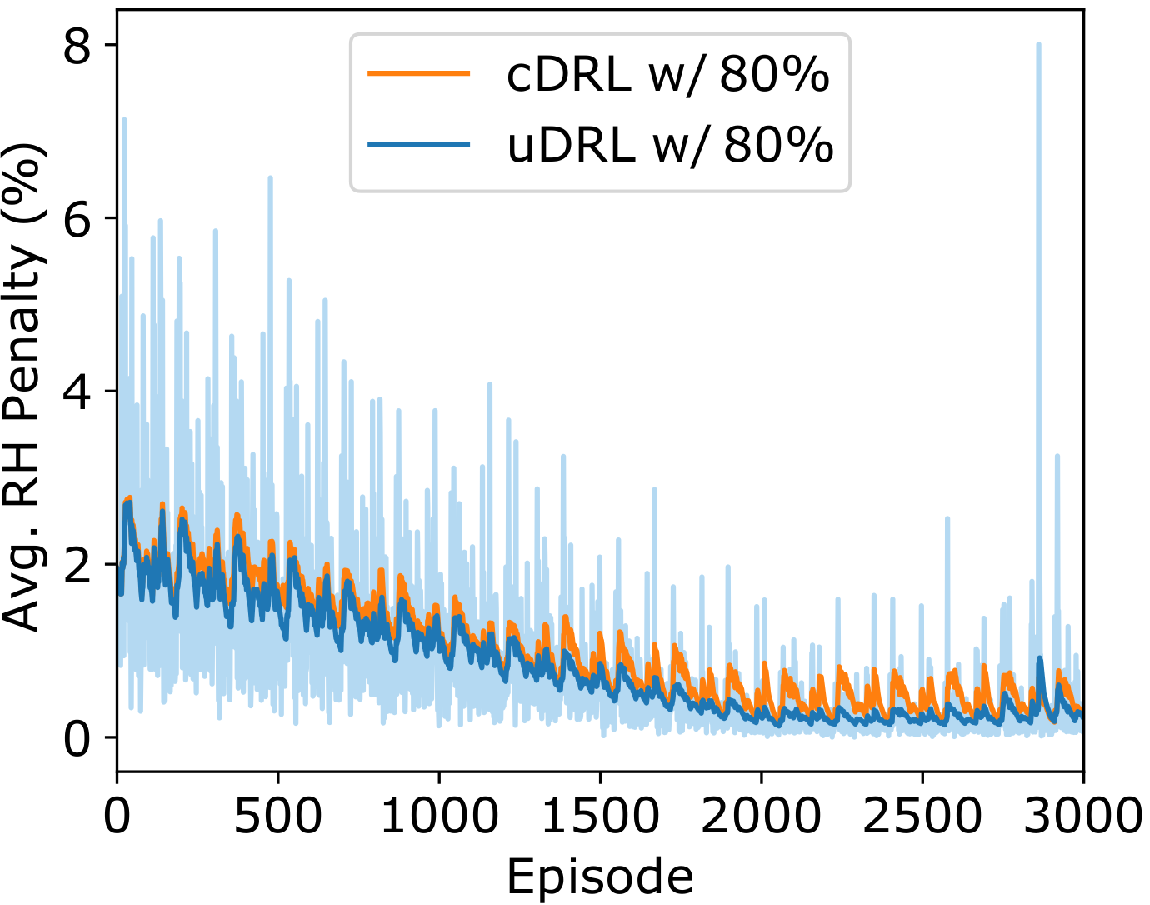}
    \caption{RH penalty.}
    \label{fig:training_rh_80}
  \end{subfigure}
  \begin{subfigure}{.32\textwidth}
    \includegraphics[width=\textwidth]{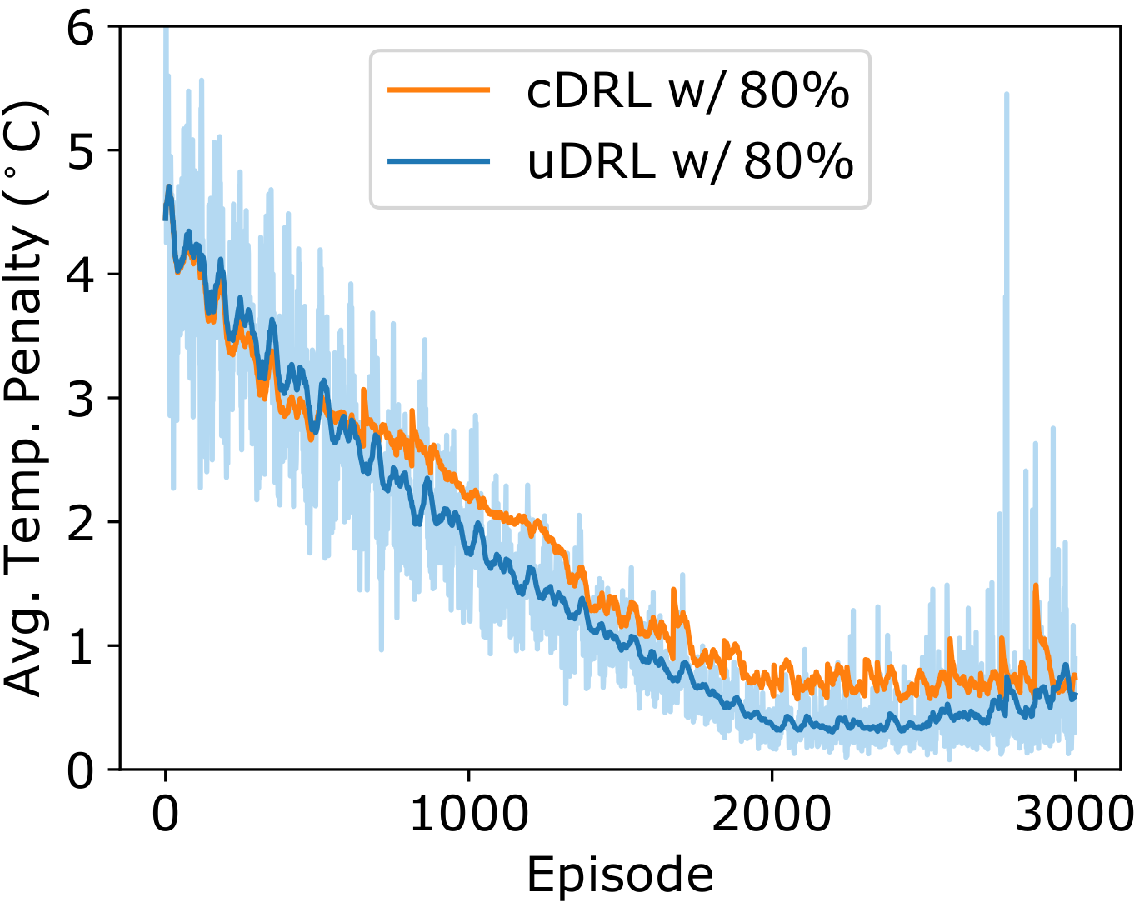}
    \caption{Temperature penalty.}
    \label{fig:training_temp_80}
  \end{subfigure}
  \begin{subfigure}{.32\textwidth}
    \includegraphics[width=\textwidth]{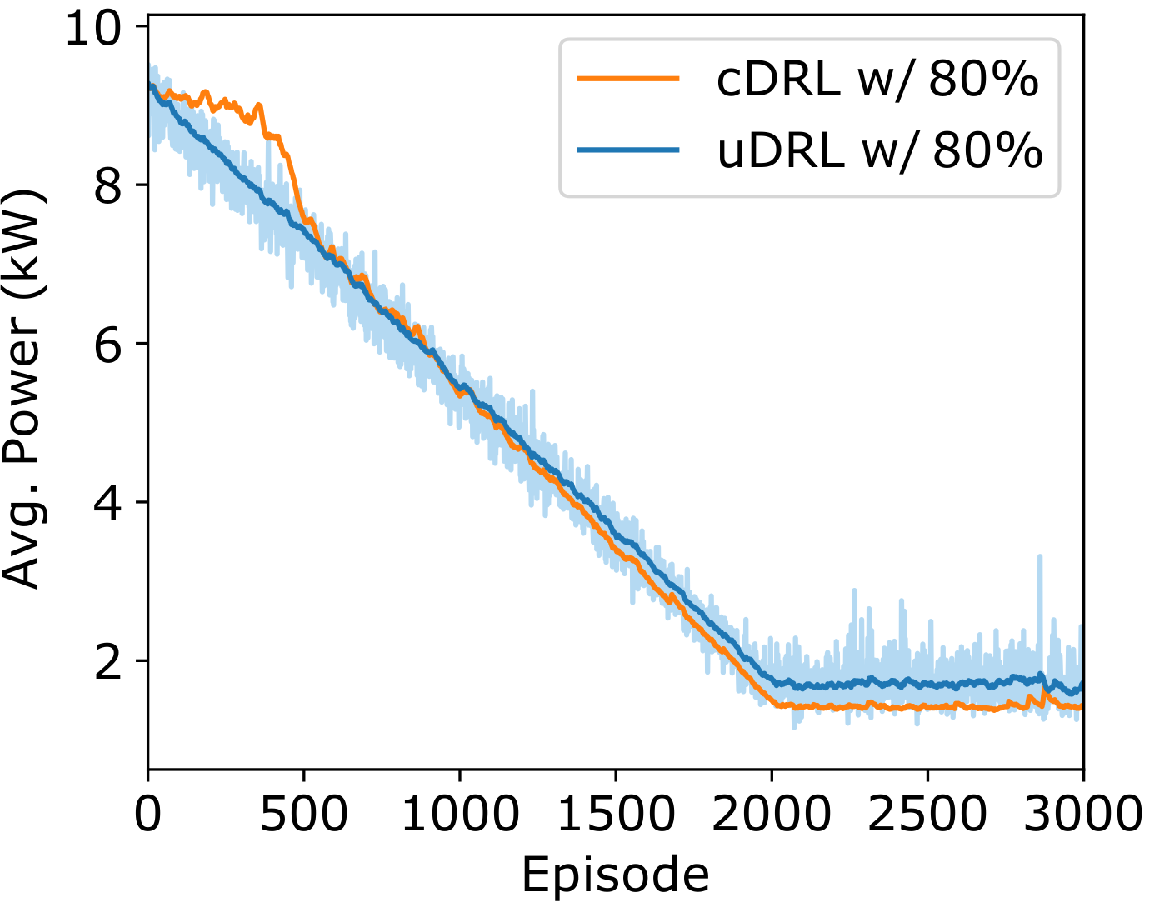}
    \caption{Cooling power.}
    \label{fig:training_energy_80}
  \end{subfigure}
  \caption{Training performance with $t_{\text{th}}$ = 32\textdegree{}C and $\phi_{\text{th}} = 80\%$.}
  \label{fig:training_80}
\end{figure}

The cDRL converges to a more stable policy with lower variance of the power, temperature and RH penalties.
More specifically, as shown in Fig.~\ref{fig:training_65}, with $t_{th}$ = 32\textdegree{}C and $\phi_{th} = 65\%$, the RH penalty of cDRL converges to lower values than those of the uDRL.
The temperature penalties of the two approaches have similar trends during the training.
Moreover, the cDRL always exhibits a higher cooling power than that of the uDRL during the training.
This is because the cDRL converges to a control policy that recirculates more hot return air and
mixes it with the outside air to maintain to the lower supply RH $\phi_s$ as shown in Fig.~\ref{fig:training_rh_65}.
When a larger amount of the hot return air is recirculated, the cooling coil is required to provide
a higher cooling capability (i.e., higher temperature reduction $\Delta t$)
to maintain the low supply air temperature $t_s$.
From Fig.~\ref{fig:training_80}, when the supply air RH requirement $\phi_{th}$ is relaxed to the higher threshold $\phi_{th} = 80\%$, both cDRL and uDRL agents can converge to a more stable policy
with lower RH and temperature penalties and cooling power.
The RH and temperature penalties are close to zero at the end of the training.

\subsubsection{DRL agent execution}

In this section, we evaluate the execution of the trained cDRL and uDRL agents for controlling the system in trace-driven simulations over a period of 5,000 time steps (i.e., about 3.5 days).
The last 3.5 days' outdoor air condition trace shown in Fig.~\ref{fig:outside_condition} is used to drive the simulations. We compare our DRL-based approaches with a hysteresis-based approach. The hysteresis-based approach adopts the maximum setpoints for the fans, i.e., $\dot{v}_s = 10,000\text{m}^3/\text{h}$.
The initial setpoints of the cooling coil $\Delta t$\textdegree{}C and damper system $\alpha$ are set to 15\textdegree{}C and 1. At the beginning of every control period, if the current supply air RH $\phi_s < \phi_{\text{th}}$,
the $\alpha$ is decreased by 0.1; otherwise, $\alpha$ is increased by 0.1.
The $\Delta t$ is decreased by 1\textdegree{}C if the current supply temperature $t_s < t_{th}$.
Otherwise, when $t_s > t_{th}$, the $\Delta t$ is increased by 1\textdegree{}C.
In the case that the $\alpha$ is already set to its maximum value of 1 but the $\phi_s$ is still higher than the threshold $\phi_{\text{th}}$, the $\Delta t$ is decreased by 1\textdegree{}C if the $t_s < t_{th}$.
This is because with the $\Delta t > 0$, the cooling coil not only reduces the supply temperature
but also increases the supply RH. The cooling coil does not change the moisture content of its processed air.
However, the cooled air leaving the cooling coil has a higher RH due to the reduced air temperature at the same moisture content. Therefore, the hysteresis-based approach decreases the $\Delta t$ to reduce the supply RH if the $t_s < t_{th}$.

\begin{figure}
  \centering
  \begin{subfigure}{.325\textwidth}
    \includegraphics[width=\textwidth]{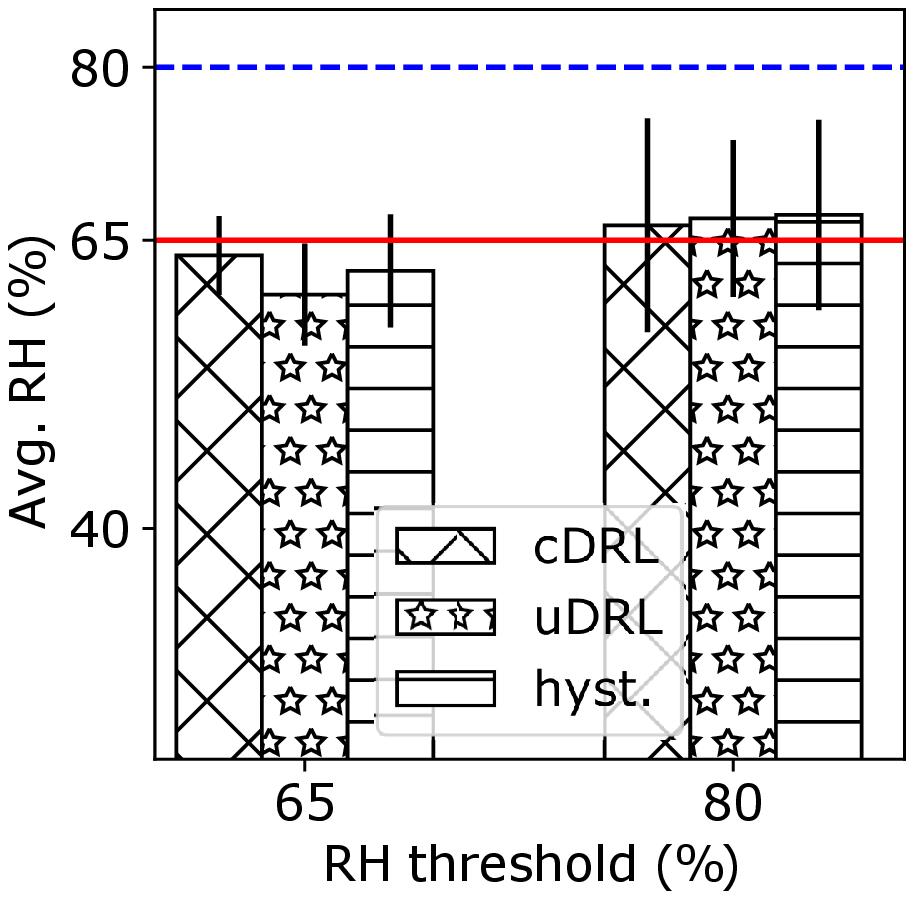}
    \caption{$t_{\text{th}}$ = 32\textdegree{}C.}
    \label{fig:testing_RH_32}
  \end{subfigure}
  \begin{subfigure}{.325\textwidth}
    \includegraphics[width=\textwidth]{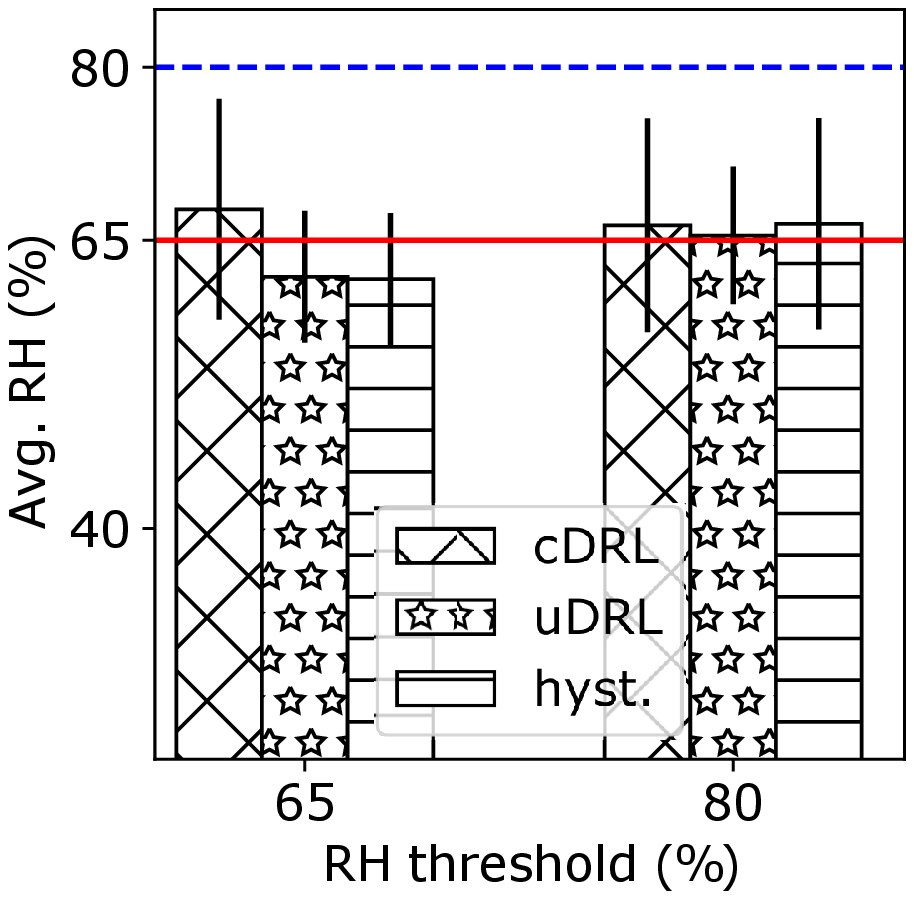}
    \caption{$t_{\text{th}}$ = 35\textdegree{}C.}
    \label{fig:testing_RH_35}
  \end{subfigure}
  \begin{subfigure}{.325\textwidth}
    \includegraphics[width=\textwidth]{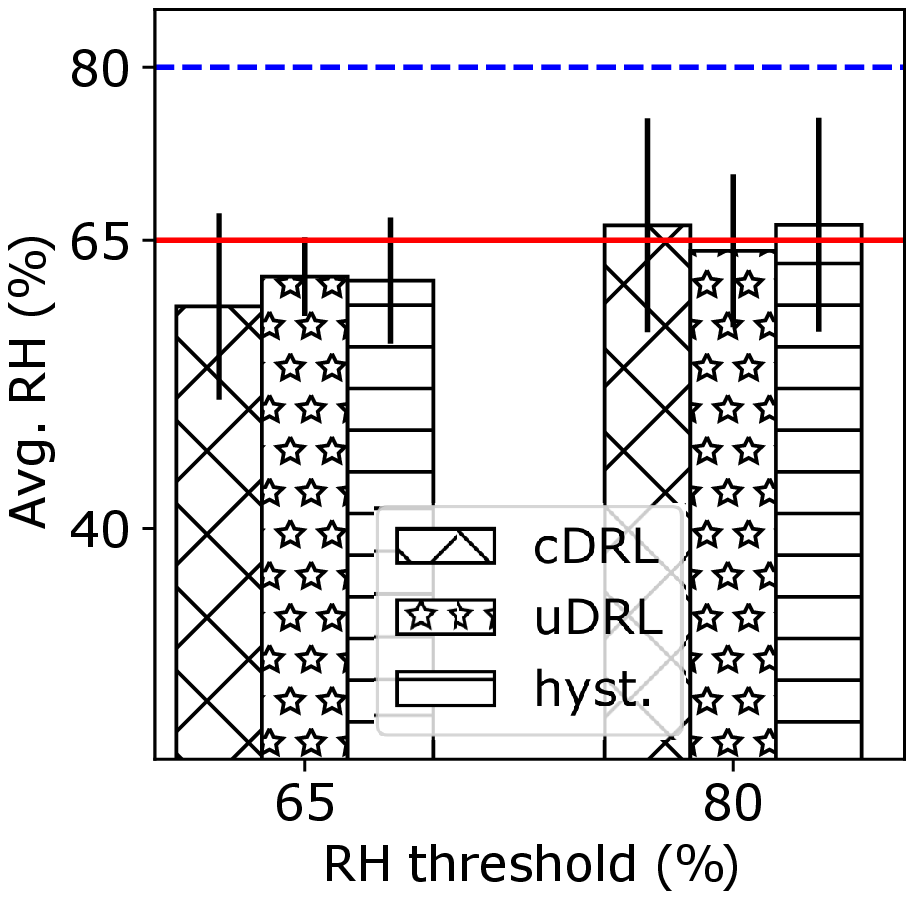}
    \caption{$t_{\text{th}}$ = 40\textdegree{}C.}
    \label{fig:testing_RH_40}
  \end{subfigure}
  \caption{Execution RH results.
  The red and dotted blue lines in subfigures represent $\phi_{th}$ of 65\% and
  80\%, respectively.}
  \label{fig:testing_rh}
\end{figure}

In the preliminary version of this work~\cite{le:2019}, we compared our uDRL approach with the model predictive control (MPC) approach. Specifically, the MPC controller schedules 10 future actions such that the predicted total cooling energy consumption is minimized subject to the temperature and RH constraints (i.e., $t_{\text{th}}$ and $\phi_{\text{th}}$).
Our preliminary results show the MPC can find the optimal policy that minimizes the cooling power,
while satisfying the supply air temperature and RH constraints.
However, the MPC requires extensive computation which leads to a long execution time.
For example, on a workstation computer with a $3.5\,\text{GHz}$ CPU and $16\,\text{GB}$ RAM,
the MPC needs 572.67 seconds on average to determine an action while our uDRL needs only 0.014 seconds.
Since each control period is 60 seconds only, the MPC approach violates the timeliness requirement.
Thus, the MPC cannot be applied in practice. Moreover, the excessive time for executing the MPC solver prevents us from running the simulations for long simulated time. Therefore, in this work, we focus on compare our new cDRL approach with the uDRL and hysteresis-based approaches only. The detailed results of the MPC approach can be found in our preliminary work~\cite{le:2019}.

Fig.~\ref{fig:testing_rh} shows the average supply RH over the execution period of 7 days under the control polices learned by the cDRL, uDRL and hysteresis-based approaches when the $t_{th}$ and $\phi_{th}$ are varied in ranges of [32\textdegree{}C, 35\textdegree{}C, 40\textdegree{}C] and [65\%, 80\%], respectively. The average supply RHs of three cDRL, uDRL and hysteresis-based approaches are always less than the $\phi_{th}$.
These results imply that three approaches can always find a control policy that maintains the average supply RH
below the required threshold $\phi_{th}$.
Moreover, the cDRL approach mostly has the lowest supply RH variance than that of the uDRL approach.

\begin{figure}
  \centering
  \begin{subfigure}{.325\textwidth}
    \includegraphics[width=\textwidth]{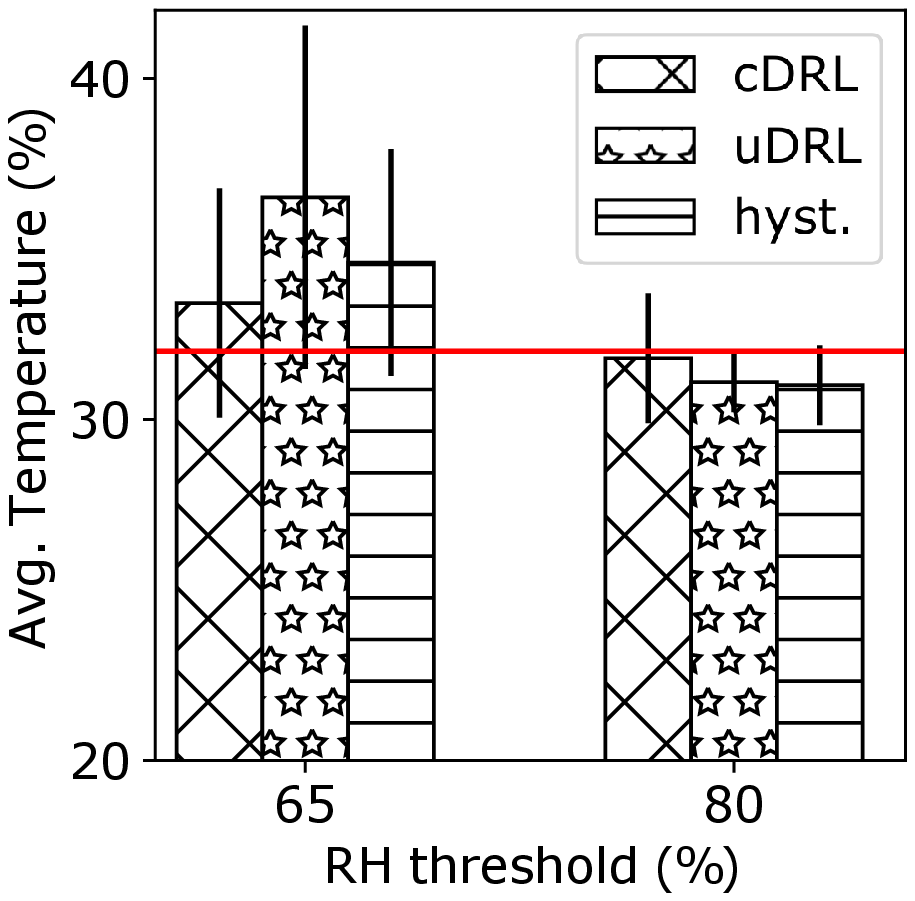}
    \caption{$t_{\text{th}}$ = 32\textdegree{}C.}
    \label{fig:testing_temp_32}
  \end{subfigure}
  \begin{subfigure}{.325\textwidth}
    \includegraphics[width=\textwidth]{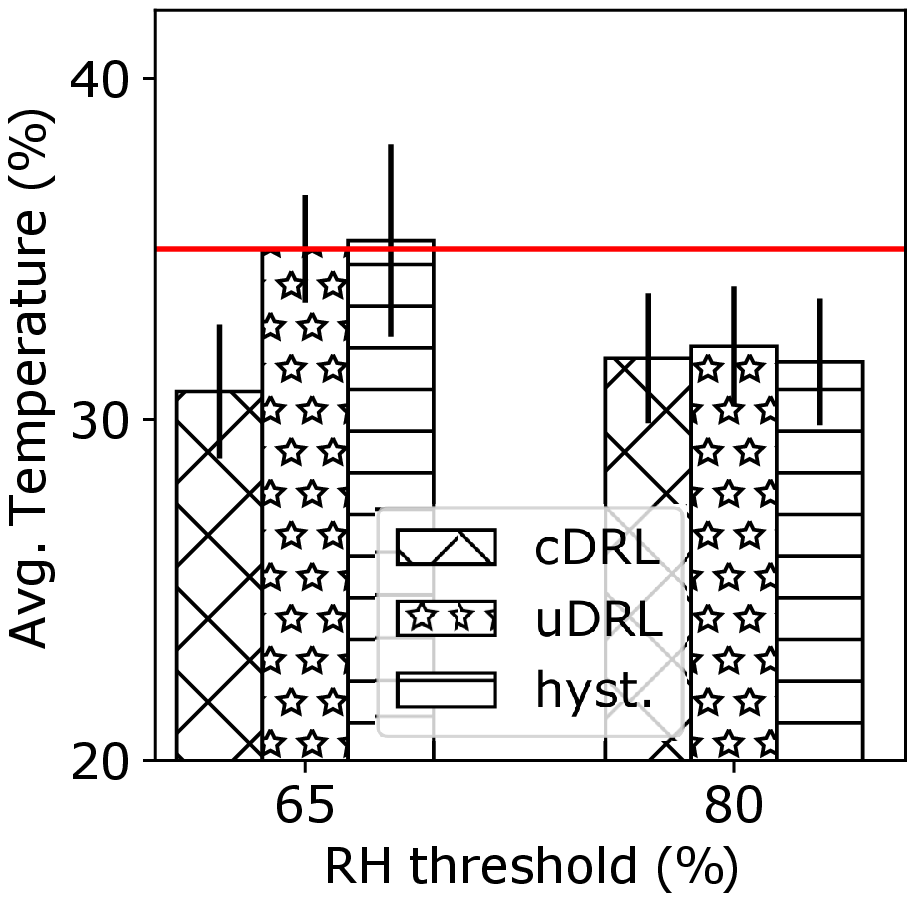}
    \caption{$t_{\text{th}}$ = 35\textdegree{}C.}
    \label{fig:testing_temp_35}
  \end{subfigure}
  \begin{subfigure}{.325\textwidth}
    \includegraphics[width=\textwidth]{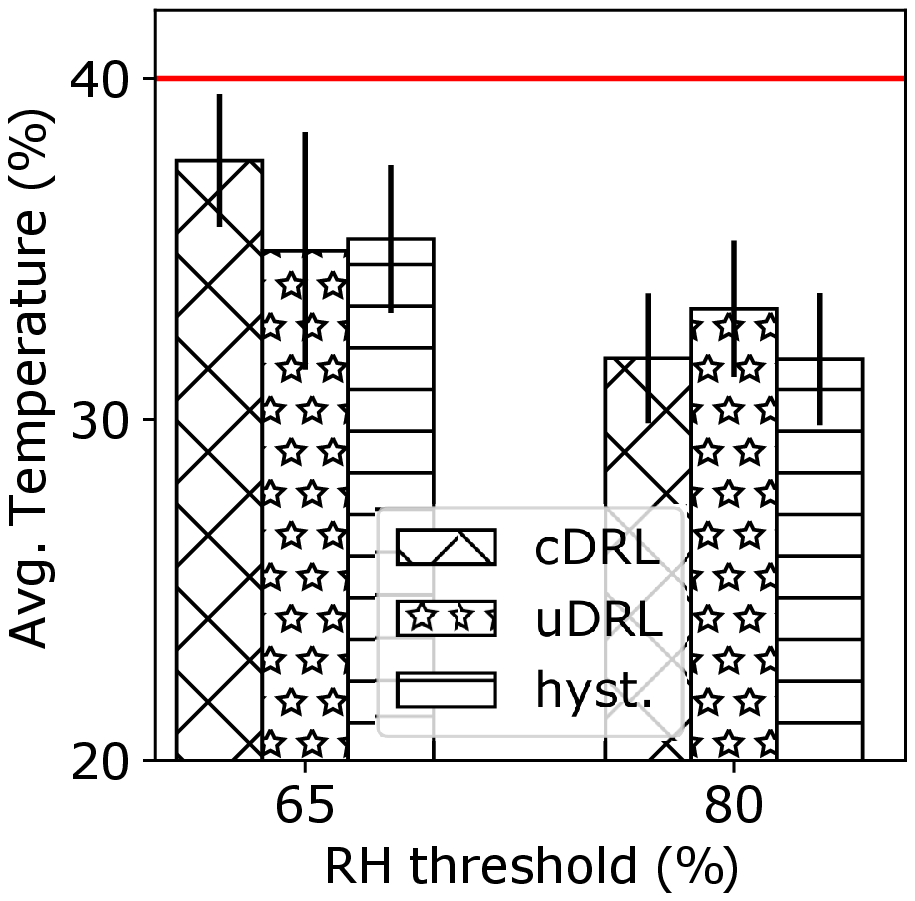}
    \caption{$t_{\text{th}}$ = 40\textdegree{}C.}
    \label{fig:testing_temp_40}
  \end{subfigure}
  \caption{Execution temperature results. The red lines in three subfigures represent $t_{th}$ of 32\textdegree{}C,
  35\textdegree{}C and 40\textdegree{}C, respectively.}
  \label{fig:testing_temp}
\end{figure}

\begin{figure}
  \centering
  \begin{subfigure}{.325\textwidth}
    \includegraphics[width=\textwidth]{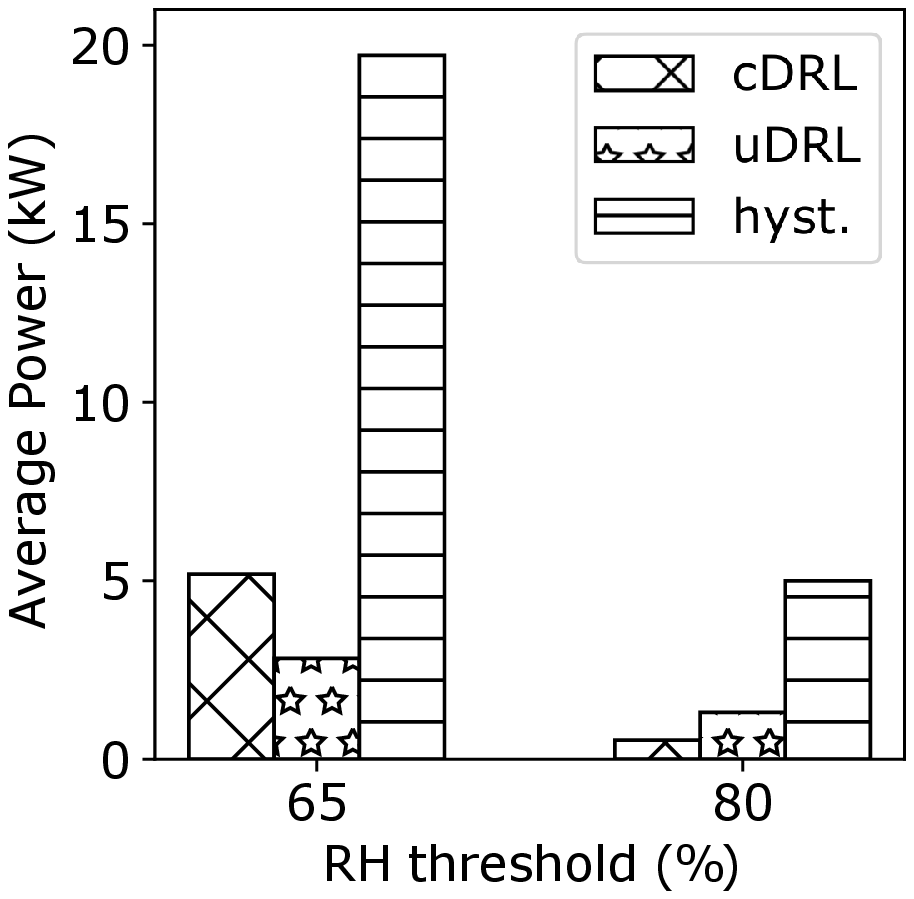}
    \caption{$t_{\text{th}}$ = 32\textdegree{}C.}
    \label{fig:testing_energy_32}
  \end{subfigure}
  \begin{subfigure}{.325\textwidth}
    \includegraphics[width=\textwidth]{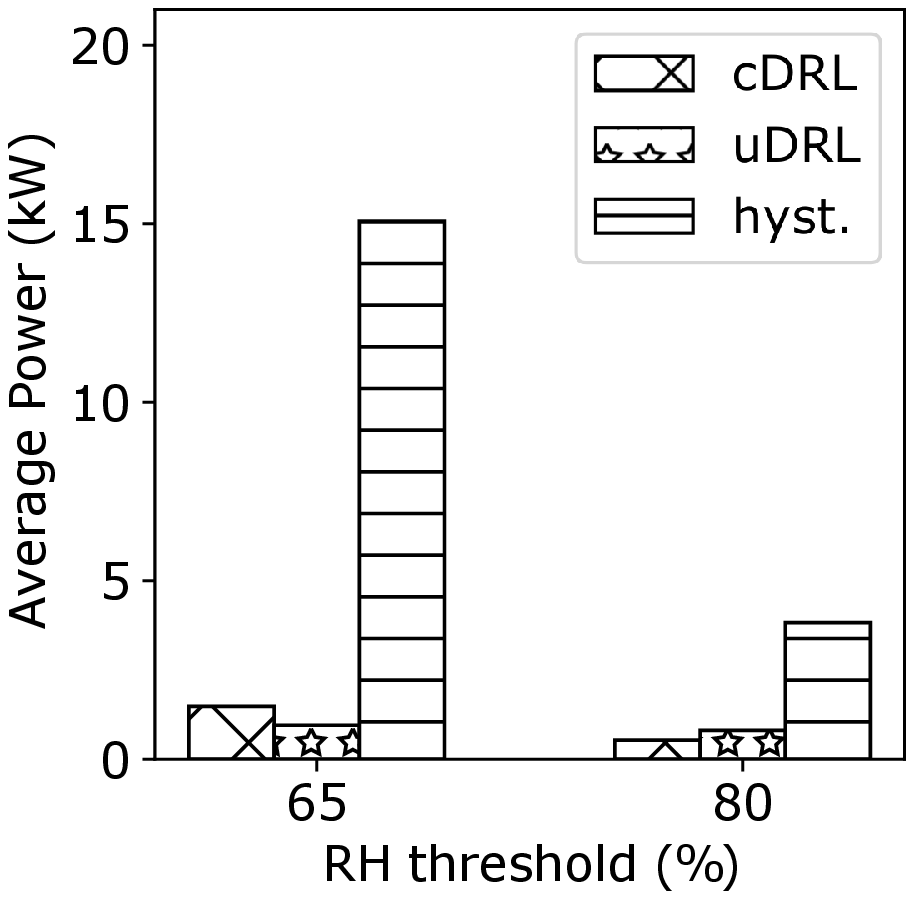}
    \caption{$t_{\text{th}}$ = 35\textdegree{}C.}
    \label{fig:testing_energy_35}
  \end{subfigure}
  \begin{subfigure}{.325\textwidth}
    \includegraphics[width=\textwidth]{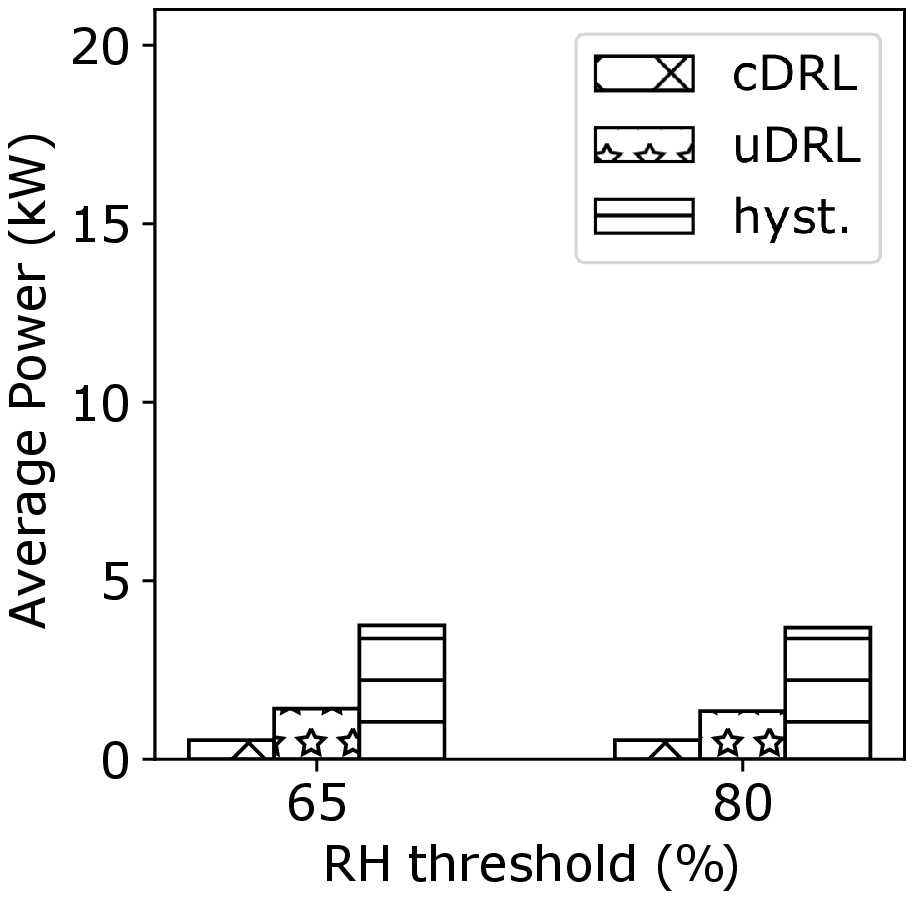}
    \caption{$t_{\text{th}}$ = 40\textdegree{}C.}
    \label{fig:testing_engergy_40}
  \end{subfigure}
  \caption{Execution cooling power results.}
  \label{fig:testing_energy}
\end{figure}

Fig.~\ref{fig:testing_temp} presents the average supply temperature.
Under all settings of ($t_{th}$ and $\phi_{th}$), except (32\textdegree{}C and 65\%),
three control approaches can maintain the average temperatures below the required thresholds $t_{th}$.
The uDRL approach also has higher temperature variance than that of the cDRL.
Under the stringent requirement of $t_{th}$ =  32\textdegree{}C and $\phi_{th}$ = 65\%,
the three control approaches can find control policies that can satisfy the RH requirement only,
and the cDRL can achieve the lowest average temperature, compared with others.
Fig.~\ref{fig:testing_energy} shows the average cooling powers of the three approaches.
The cooling powers of the three approaches increase with the decreased $t_{th}$ and $\phi_{th}$,
since lower cooling demand and fan speed are needed.
Moreover, the hysteresis-based approach always has a higher power
than those of the cDRL and uDRL approaches. When $t_{th}$ > 35\textdegree{}C or $\phi_{th}$ > 65\%,
the cDRL has lower cooling power.

In summary, the three evaluated control approaches mostly maintain the supply air temperature and RH below the required thresholds. The hysteresis-based approach always consume more cooling energy than our uDRL and cDRL approaches. Moreover, the cDRL approach outperforms the uDRL and hysteresis-based approaches in most cases
under various settings of supply temperature and RH requirements.
Specifically, compared with the uDRL approach, the cDRL is able to find a more stable control policy with less RH and temperature requirement violations. Moreover, when the temperature and RH constraints are relaxed, the cDRL approach has the lowest cooling power consumption while still maintaining the supply temperature and RH below the required thresholds.

Furthermore, the cooling energy savings achieved by the two DRL approaches compared with the hysteresis-based approach
decrease with the $t_{th}$ and $\phi_{th}$.
In particular, the hysteresis-based approach can work well enough if the outside air temperature and RH
are always below the $t_{th}$ and $\phi_{th}$, respectively,
since only fans are used to remove the heat generated by the servers.
However, as discussed in \sect\ref{subsec:temp-requirement} and \sect\ref{subsec:RH-requirement},
the ASHRAE's relaxed temperature and RH requirements up to 45\textdegree{}C and 90\%
are for traditional DCs with clean air that is recirculated within the enclosed DC buildings only.
The outside air inhaled into the air free-cooled DCs may contain corrosive gaseous and particulate contaminants
that can cause corrosion on the IT equipment if the supply air RH is higher than the deliquescent RH
of the particles and gases (e.g., 65\%).
Moreover, the high supply air temperature can accelerate the corrosion process.
The air free-cooled DCs should adopt $t_{th}$ setting lower than the server's maximum allowed temperature.
Therefore, our proposed DRL approaches are promising for efficiently controlling the supply air condition
to meet the tight RH and temperature requirements, especially in the tropics.

\section{Conclusion}
\label{sec:conclude}

This paper developed an essential function for operating air free-cooled DCs in tropics -- the control of the temperature and RH of the air supplied to the servers. It is based on an energy-efficient design of recirculating a controlled portion of return hot air to mix with the fresh outside air.
We formulated the control problem and proposed both unconstrained and constrained DRL-based solutions.
Trace-driven simulations showed the effectiveness of our solutions.

\begin{acks}
This project is a collaboration between Info-communications Media Development Authority (IMDA) and Nanyang Technological University. This project is supported in part by the National Research Foundation (NRF), Prime Minister’s Office, Singapore under the Green Data Centre Programme (GDCP) administrated by IMDA and under the Energy Programme (EP) administrated by the Energy Market Authority (Award No. NRF2017EWT-EP003-023). Y. Wen's work is additionally supported in part by NRF under the Green Data Centre Research (GDCR) program administered by IMDA and under the Behavioural Studies in the Energy, Water, Waste and Transportation Sectors (BSEWWT) program (Award No. BSEWWT2017\_2\_06), and in part by Nanyang Technological University through the Data Science and Artificial Intelligence Research Centre (DASIR). Any opinions, findings and conclusions or recommendations expressed in this material are those of the author(s) and do not reflect the views of the funders.
\end{acks}

\bibliographystyle{ACM-Reference-Format}
\bibliography{bibliography}

\end{document}